\documentclass{revtex4}

\usepackage{amsmath,amsthm}
\usepackage{rotating}
\usepackage{graphicx}

\usepackage{amsfonts}
\newtheorem{theorem}{Theorem}[section]
\newtheorem{lemma}[theorem]{Lemma}
\newtheorem{proposition}[theorem]{Proposition}
\newtheorem{cor}[theorem]{Corollary}

\theoremstyle{remark}

\theoremstyle{definition}
\newtheorem{definition}[theorem]{Definition}

\theoremstyle{example}
\newtheorem{example}[theorem]{Example}

\theoremstyle{notation}

\newcommand{\bra}[1]{\langle#1|}
\newcommand{\ket}[1]{|#1\rangle}

\begin{document}

\title{Ultra-quantum coherent states  in a single finite quantum system }            
\author{A. Vourdas}
\affiliation{Department of Computer Science,\\
University of Bradford, \\
Bradford BD7 1DP, United Kingdom\\a.vourdas@bradford.ac.uk}

\begin{abstract}
A set of $n$ coherent states is introduced in a quantum system with $d$-dimensional Hilbert space $H(d)$.
It is shown that they resolve the identity, and also have a discrete isotropy property.
A finite cyclic group acts on the set of these coherent states, and partitions it into orbits.
A $n$-tuple representation of arbitrary states in $H(d)$, analogous to the Bargmann representation, is defined.
There are two other important properties of these coherent states which make them `ultra-quantum'. 
The first property is related to the Grothendieck formalism which studies the `edge' of the Hilbert space and quantum formalisms.
Roughly speaking the Grothendieck theorem considers a `classical' quadratic form ${\mathfrak C}$ that uses complex numbers in the unit disc,
and a `quantum' quadratic form ${\mathfrak Q}$   that uses vectors in the unit ball of the Hilbert space.
It shows that if ${\mathfrak C}\le 1$, the corresponding ${\mathfrak Q}$ might take values greater than $1$, up to the complex Grothendieck constant $k_G$.
${\mathfrak Q}$ related to these coherent states is shown to take values in the `Grothendieck region' $(1,k_G)$, which is classically forbidden in the sense that
 ${\mathfrak C}$ does not take values in it. 
The second property complements this, showing that these coherent states violate logical Bell-like inequalities (which for a single quantum system are quantum versions of the Frechet probabilistic inequalities).
In this sense also, our coherent states are deep into the quantum region.
\end{abstract}
\maketitle

\section{Introduction}
Coherent states \cite{COH1,COH2} have been studied for a long time (in relation to the Heisenberg-Weyl group, SU(2) group, SU(1,1) group, etc), but we emphasise from the outset  that our ultra-quantum coherent states are different.
In fact coherent states are sometimes described as quasi-classical states (e.g., they minimise the uncertainty relations), in contrast to our coherent states which are in the opposite `ultra-quantum' end of the spectrum.

The present work is in the context of a single quantum system with a $d$-dimensional Hilbert space $H(d)$.
Coherent states related to the Heisenberg-Weyl group in this context have been studied in refs\cite{COH3,COH4}, and are not related to the present work.
The coherent states introduced in this paper have the following properties:
\begin{itemize}
\item
{\bf C1:} Resolution of the identity.
\item
{\bf C2}: The set ${\cal C}(d,n,z)$ of $n$ coherent states in $H(d)$ (that depend on a parameter $z$ with $|z|=1$), is invariant under transformations in a finite cyclic group ${\cal G}_d$.
 The action of ${\cal G}_d$ on the set ${\cal C}(d,n,z)$ of coherent states is not transitive, and partitions it into orbits which are described with `orbit matrices'.
The orbit matrices are proportional to the matrices of overlaps of the coherent states, and are important for property {\bf C6} below.
Three sets ${\cal C}(3,6,z)$, ${\cal C}(4,8,z)$, ${\cal C}(4,12,z)$ of these coherent states, are given as examples.

\item
{\bf C3:} Discrete isotropy.
\item
{\bf C4:} A projector related to overlaps of the coherent states is the `reproducing kernel' \cite{RC} in a discrete context.
A $n$-tuple representation of states in $H(d)$ is the discrete analogue of the Bargmann analytic representation\cite{BA}. In the present discrete context
there is no analyticity, but we note that there is work on discrete versions of analyticity (e.g.,\cite{DA, DA1}) which might be useful in the present context.
The discrete isotropy leads to interesting relations among the matrix elements of this projector.
\end{itemize}

These and some other related properties are discussed in sections \ref{sec31}, \ref{sec41} and justify the use of the term coherent states.
But there are two more properties of these coherent states, which justify the term `ultra-quantum'. The first is related to the Grothendieck formalism which studies the `edge' of the Hilbert space and quantum formalism, and justifies the term `ultra-quantum'.
The second complements this, showing that our coherent states violate Bell-like inequalities, and in this sense are far from quasi-classical behaviour and deep into the quantum region.
They are described briefly below:
\begin{itemize}

\item
{\bf C5: The Grothendieck quadratic expression ${\mathfrak Q}$ with our coherent states, takes  values in the `Grothendieck region' $(1,k_G)$:}
Roughly speaking the Grothendieck formalism considers a `classical' quadratic form ${\mathfrak C}$ that uses complex numbers in the unit disc, and takes values less than $1$.
It then proves that if the complex numbers are replaced with vectors in the unit ball of the Hilbert space, then the `quantum' quadratic form ${\mathfrak Q}$   might take values greater than $1$, up to the complex Grothendieck constant $k_G$.
Going from  ${\mathfrak C}$ to ${\mathfrak Q}$ replaces scalars with vectors, and is a passage from classical to quantum mechanics.
The `Grothendieck region' $(1,k_G)$ is of special interest, because it is a classically forbidden region (${\mathfrak C}$ cannot take values in it).

The original formulation of the Grothendieck theorem \cite{QR1} was in the context of tensor product of Banach spaces,
and this leads to the impression that applications in a quantum context should be for multipartite systems described by tensor products of Hilbert spaces.
Indeed, in a quantum context the  Grothendieck formalism has been applied to multipartite entangled systems \cite{QR2,QR3,QR4,QR5,QR6,QR7,QR8,QR9}, and violation of Bell-like inequalities
has been linked to the fact that  ${\mathfrak Q}$ can take values greater than one.
Later mathematical work on Grothendieck's theorem \cite{AR1,AR2,AR3} emphasised that the theory does not require tensor products.

In a quantum context ref\cite{VOU1} used the Grothendieck formalism for a single quantum system.
The present work expands significantly this direction, with a deeper understanding of the properties {\bf C1-C4}, and more examples (e.g., the set ${\cal C}(4,8,z)$) of coherent states that lead to 
${\mathfrak Q}$ with  values in the Grothendieck region $(1,k_G)$. A new direction is to show that these coherent states violate  logical Bell-like inequalities, as discussed in the property {\bf C6} below.

\item
{\bf C6: Our coherent states violate  logical Bell-like inequalities for a single quantum system:}
Logical Bell-like inequalities have been introduced in ref.\cite{AH} in the context of multipartite entangled systems, and used in refs.\cite{AH1,AH2}. 
In the context of multipartite systems they describe correlations between the various parties.
In a single quantum system we have no such correlations, and they resemble the Frechet probabilistic inequalities which are known in  a classical context.
But we note that their logical derivation (proposed in ref\cite{AH}), is the same for a single system or for multipartite systems.
In the present paper their violation by  projectors related to our coherent states in a single quantum system, is interpreted as an indication of their `quantumness'.

Quantum Mechanics violates these inequalities because their proof is based on Boole's inequality which holds for Kolmogorov probabilities and is violated by quantum probabilities.
In our presentation of Bell-like inequalities, we emphasise the assumptions used in their derivation.
Their violation is a natural consequence of the fact that these assumptions do not hold.
In section \ref{sec57} we use the orbit matrices to show that projectors related to the coherent states in any of the orbits, violate 
logical Bell-like inequalities.  Therefore our coherent states are far from quasi-classical behaviour, and they are deep into the quantum region.

\end{itemize}

In section 2 we describe a quantum system with $d$-dimensional Hilbert space $H(d)$, and show how the action of the finite cyclic group ${\cal G}_d$ on its states,  leads to orbits.
We also introduce circulant density matrices that describe these orbits.
In section 3 we introduce our coherent states.
We first summarise briefly their properties, and then in subsections \ref{sec1},\ref{sec2},\ref{sec3} we give three families of examples and discuss in detail these properties.
In section 4 we present a $n$-tuple representation of arbitrary states in $H(d)$, analogous to the Bargmann representation.

Sections 5,6 present the `ultra quantum' properties of our coherent states.
Section 5 shows that ${\mathfrak Q}$ related to our coherent states takes values in the `Grothendieck region' $(1,k_G)$, which is classically forbidden in the sense that
 ${\mathfrak C}$ does not take values in it.  We note that large families of physically interesting  quantities lead to ${\mathfrak Q}\le 1$ (example \ref{ex1}), and in this sense 
 examples with ${\mathfrak Q}> 1$ are rare.
 
  Section 6, shows that logical Bell-like inequalities for a single quantum system, are violated by our coherent states. 
 We conclude in section 7 with a discussion of our results.

\section{Preliminaries}

We consider a quantum system with variables in ${\mathbb Z}_d$ (the ring of integers modulo $d$) and a $d$-dimensional Hilbert space $H(d)$.
Let $\ket{X;r}$ with $r\in {\mathbb Z}_d$ be an orthonormal basis which we call position states ($X$ in this notation is not a variable, and indicates `position states').
All states and operators below are written as vectors and matrices in the position basis.

 Let $h_1,h_2$ be
subspaces of $H(d)$, and ${\cal O}$ its zero-dimensional subspace that contains only the zero vector. We denote as 
$\prec$ the partial order `subspace'. For later use (in section 6) we define the conjunction ($\wedge$), disjunction ($\vee$) and negation ($\neg$) of subspaces as
\begin{eqnarray}
h_1\wedge h_2=h_2\cap h_1;\;\;\;h_1\vee h_2={\rm span}(h_1\cup h_2);\;\;\;\neg h_1=h_1^\perp.
\end{eqnarray} 
We note that the disjunction is not just the union of the two spaces, but it also contains all superpositions of vectors in the two spaces.
Both disjunction and conjunction can be used with a sequence of many subspaces.
The negation is the orthocomplement $h_1^\perp$of $h_1$, and has the properties:
\begin{eqnarray}
&&h_1\wedge h_1^\perp={\cal O};\;\;\;h_1\vee h_1^\perp=H(d);\;\;\;(h_1^\perp)^\perp=h_1\nonumber\\
&&(h_1\wedge h_2)^\perp=h_1^\perp \vee h_2^\perp;\;\;\;(h_1\wedge h_2)^\perp=h_1^\perp \vee h_2^\perp.
\end{eqnarray}
Also
\begin{eqnarray}\label{PP}
&&\Pi(h_1)+\Pi(h_1^\perp)={\bf 1}_d.
\end{eqnarray} 
Here $\Pi(h_1)$ is the projector to the subspace $h_1$ and ${\bf 1}_d$ is the $d\times d$ identity matrix.

The set ${\cal L}[H(d)]$ of subspaces of $H(d)$ with the above operations is a modular orthocomplemented lattice (e.g. \cite{L1,L2,L3} or chapter 6 in \cite{COH4}).
The conjunction is the intersection of $h_1,h_2$ as in Boolean algebra and set theory. 
Disjunction is here very different from Boolean algebra and set theory because it is not just the union $h_1\cup h_2$, but includes all superpositions.
This is an important difference between classical physics which is based on Boolean logic, and quantum physics with finite dimensional Hilbert space described by 
a modular orthocomplemented lattice.

\subsection{Orbits in $H(d)$ generated by the finite cyclic group ${\cal G}_d$}
We represent the position states $\ket{X;s}$ with columns which have zeros everywhere except the $s$ row where they have one:
\begin{eqnarray}
\ket{X;s}=(...\delta_{s\nu}....)^T.
\end{eqnarray}
Let $X$ be the $d\times d$ `upwards displacement' matrix (circulant permutation matrix):
\begin{eqnarray}\label{1X}
X=\begin{pmatrix}
0&1&0&\cdots&0\\
0&0&1&\cdots&0\\
\vdots&\vdots&\vdots&\ddots&\vdots\\
0&0&0&\cdots&1\\
1&0&0&\cdots&0
\end{pmatrix};\;\;\;X^r\ket{X;s}=\ket{X;s-r};\;\;\;X^d={\bf 1}_d;\;\;\;{\bf 1}+X+X^2+...+X^{d-1}=J_d
\end{eqnarray}
Here $J_d$ is the `matrix of ones' (all elements are equal to $1$). $\frac{1}{d}J_d$ is a projector with eigenvalues $1$ (with multiplicity $1$) and $0$ (with multiplicity $d-1$).

${\cal G}_d=\{{\bf1}_d,X,...,X^{d-1}\}$ is a multiplicative group isomorphic to the additive group ${\mathbb Z}_d$.
The group ${\cal G}_d$ acts on the states in $H(d)$, and the orbit of a state $\ket{f}$ is 
\begin{eqnarray}
{\rm orb}(\ket{f})=\{X^r\ket{f}\;|\;r=0,...,d-1\},
\end{eqnarray}
and it is an equivalence class. Its cardinality is $|{\rm orb}(\ket{f})|\le d$.

We consider stroboscopic time evolution described with the operator $X^t$ where the time $t$ is an integer.
Then the state $\ket{f}$ evolves periodically to all states in orbit ${\rm orb}(\ket{f})$ with period $|{\rm orb}(\ket{f})|$.
In order to describe the `average quantum state' during the stroboscopic time evolution of $\ket{f}$, 
we add all projectors associated with the states in the orbit, and we get an operator which is Hermitian and positive semi-definite (as sum of projectors).
We divide this over the period $|{\rm orb}(\ket{f})|$ and we get the `orbit density matrix' 
\begin{eqnarray}
R=\frac{1}{|{\rm orb}(\ket{f})|}\sum _rX^r\ket{f}\bra{f}X^{-r};\;\;\;[R, X]=0.
\end{eqnarray}
It is known (e.g., \cite{CIRC}) that matrices that commute with $X$ are circulant matrices, and therefore 
\begin{eqnarray}
R=\frac{1}{d}{\bf 1}_d+r_1X+...+r_{d-1}X^{d-1};\;\;\;r_k=r_{d-k}^*.
\end{eqnarray}
Since $R$ is Hermitian matrix it follows that $r_k=r_{d-k}^*$ (and for even $d$ the $r_{d/2}$ is real).
Circulant matrices commute with each other, and therefore all orbit density matrices commute with each other.
The eigenvectors of $d\times d$ circulant matrices are known (e.g., \cite{CIRC}) to be the columns of the $d\times d$ Fourier transform
\begin{eqnarray}
F_{\mu \nu}=\frac{1}{\sqrt{d}}\omega^{\mu\nu};\;\;\;\omega=\exp\left(i\frac{2\pi}{d}\right);\;\;\;\mu,\nu\in {\mathbb Z}_d.
\end{eqnarray}
Therefore
\begin{eqnarray}
R\ket{{\mathfrak f}_\nu}=\left (\frac{1}{d}+r_1\omega^\nu+...+r_{d-1}\omega^{\nu(d-1)}\right)\ket{{\mathfrak f}_\nu}
;\;\;\;
\ket{{\mathfrak f}_\nu}=\frac{1}{\sqrt{d}}
\begin{pmatrix}
1\\\omega ^\nu\\\vdots\\\omega^{\nu(d-1)}
\end{pmatrix}.
\end{eqnarray}
Below we generalise the orbit density matrices  into orbit matrices that link different orbits.

The orbit density matrix describes the average quantum state in the orbit.
The average expectation value of an operator $\theta$ during the stroboscopic time evolution of $\ket{f}$, is nicely expressed in terms of the orbit density matrix as
\begin{eqnarray}
\frac{1}{|{\rm orb}(\ket{f})|}\sum _r\bra{f}X^{-r}\theta X^r\ket{f}={\rm Tr}(R \theta).
\end{eqnarray}

We note here that orbits and other related group theory methods have been used in the analysis of quantum systems in refs\cite{OR1,OR2,OR3,OR4}.

\begin{example}
We consider the group ${\cal G}_3$ acting on $H(3)$.
Examples of orbits of states (expressed in the position basis)  are:
\begin{eqnarray}
{\rm orb}\left (\frac{1}{\sqrt{3}}
\begin{pmatrix}
1\\1\\ 1
\end{pmatrix}\right )=\left\{\frac{1}{\sqrt{3}}
\begin{pmatrix}
1\\1\\ 1
\end{pmatrix}\right \};\;\;\;
{\rm orb}\left (\frac{1}{\sqrt{14}}
\begin{pmatrix}
1\\2\\ 3
\end{pmatrix}\right )=\left\{\frac{1}{\sqrt{14}}
\begin{pmatrix}
1\\2\\ 3
\end{pmatrix},
\frac{1}{\sqrt{14}}
\begin{pmatrix}
2\\ 3\\1
\end{pmatrix},
\frac{1}{\sqrt{14}}
\begin{pmatrix}
 3\\1\\2
\end{pmatrix}
\right \}
\end{eqnarray}
The corresponding density matrices are the circulant matrices:
\begin{eqnarray}\label{17}
R_1=\frac{1}{3}J_3;\;\;\;
R_2=\frac{1}{3}{\bf 1}_3+\frac{11}{42}(X+X^\dagger)=\frac{1}{14}{\bf 1}_3+\frac{11}{42}J_3
\end{eqnarray}

\end{example}

\section{Coherent states }\label{sec31}
Three examples given below prove the  existence of sets ${\cal C}(d,n,z)$ of $n$ (an integer multiple of $d$) normalised states $\ket{a_z(r)}$ ($r=0,...,n-1$) in $H(d)$, which we call 
coherent states because of the properties listed below. 
$z$ is a parameter with $|z|=1$, and for each value of $z$ we get a different set of coherent states.
Some of the results below are for `generic $z$' (or `almost all $z$') , which means `for all values of $z=\exp(i\theta)$ except from a finite number'. 

Below we summarise briefly their properties, which are proved and expanded later in each example.
\begin{itemize}
\item
{\bf Resolution of the identity:}
They obey the relation:
\begin{eqnarray}\label{1}
\frac{d}{n}\sum_{r=0}^{n-1}\ket{a_z(r)}\bra{a_z(r)}={\bf 1}_d.
\end{eqnarray}

\item

{\bf The action of ${\cal G}_d$ on the set of coherent states  ${\cal C}(d,n,z)$  partitions it into orbits ${\cal C}_\mu(d,n,z)$:}
Let ${\cal C}_\mu(d,n,z)$ be the set of coherent states in the $\mu$-orbit. Then
\begin{eqnarray}
{\cal C}(d,n,z)={\cal C}_0(d,n,z)\cup\cdots\cup{\cal C}_{\mathfrak N}(d,n,z);\;\;\;{\mathfrak N}=\frac{n}{d}-1.
\end{eqnarray}
The group ${\cal G}_d$ acts transitively on the states that belong to the same orbit.
The stabilisers of all coherent states are $\{{\bf 1}_d\}$ and therefore each orbit has exactly $d$ coherent states (orbit-stabiliser theorem).

In addition to the above notation, it will be convenient to introduce a `pair of indices notation'  where
\begin{eqnarray}\label{nota}
\ket{a_z(r)}=\ket{a_z({\hat r},\mu)};\;\;\;r=\hat r+\mu d;\;\;\;\hat r=r({\rm mod}\; d);\;\;\;\hat r\in{\mathbb Z}_d;\;\;\;r=1,...,n-1;\;\;\;\mu=0,...,{\mathfrak N}
\end{eqnarray}
The second index is `orbit index' and the first one (denoted with a hat) is `index within a certain orbit'.
This creates a bijective map between the indices
\begin{eqnarray}
\{0,...,n-1\}\leftrightarrow\{(0,0),...,(d-1,0),...,(0,{\mathfrak N}),...,(d-1,{\mathfrak N})\}.
\end{eqnarray}
With the `pair of indices notation' coherent states in the same equivalence class have the same second index, and
\begin{eqnarray}
\ket{a_z({\hat r},\mu)}=X^{\hat r}\ket{a_z(0,\mu)};\;\;\;\hat r\in{\mathbb Z}_d.
\end{eqnarray}
Within a given orbit, acting successively with $X$ we get all the states in the orbit (e.g., Eqs(\ref{O1}), (\ref{O2}) below).

There is  `orthogonality between the orbits', in the sense that
corresponding coherent states in different orbits are orthogonal to each other:
\begin{eqnarray}\label{Y6}
{\cal T}_{\mu \nu}(\hat r, \hat r)=\bra{a_z({\hat r},\mu)}a_z({\hat r},\nu)\rangle=\delta_{\mu\nu}.
\end{eqnarray}
This will be verified in the three examples below. This can also be written in the single index notation as
\begin{eqnarray}\label{OR}
\bra{a_z(r)}a_z(r+d\mu)\rangle=\delta_{0\mu };\;\;\;\mu=0,...,{\mathfrak N}.
\end{eqnarray}
Unlike `standard' coherent states which are non-orthogonal to each other, here we have orthogonality between 
corresponding coherent states in different orbits.

\item
{\bf For generic $z$, the coherent states in any of the orbits span $H(d)$ (but there is no resolution of the identity in terms of them only):}
If $h({\hat r},\mu)$ is the one-dimensional subspace of $H(d)$ that contains the coherent state $\ket{a_z({\hat r},\mu)}$ then
\begin{eqnarray}\label{AAD}
\bigvee _{{\hat r} =0}^{d-1} h({\hat r},\mu)=H(d).
\end{eqnarray}
Indeed in all examples below, it can be easily seen that the $d\times d$ determinant that has the coherent states $\ket{a_z({\hat r},\mu)}$ (with fixed orbit $\mu$) as columns, is non-zero (for generic $z$).
Eq.(\ref{AAD}) will be used with Bell-like inequalities, in section 6 below.

\item{\bf The orbit matrices of the coherent states are proportional to the matrices of their overlaps }
\begin{definition}
The orbit matrix that connects corresponding coherent states in the $\mu$ and $\nu$ orbits, is the following $d\times d$ 
circulant matrix in the position basis:
\begin{eqnarray}\label{ABC}
&&\sigma_{\mu\nu}(\hat s,\hat r)=\frac{1}{d}\sum_{\hat q=0}^{d-1}\langle X;\hat s \ket{a_z(\hat q,\nu)}\bra{a_z(\hat q, \mu)}X; \hat r\rangle 
=\frac{1}{d}\sum_{\hat q=0}^{d-1}\bra{X;\hat s}X^{\hat q}\ket{a_z(0,\nu)}\bra{a_z(0, \mu)}X^{-\hat q}\ket{X;\hat r}\nonumber\\
&&[X, \sigma_{\mu\nu}]=0;\;\;\;\sigma_{\mu\nu}(\hat s,\hat r)=\sigma_{\nu\mu}^\dagger(\hat s,\hat r),
\end{eqnarray}
where $\mu, \nu=0,...,{\mathfrak N}$. 
The `dagger' refers to the indices $\mu, \nu$.
\end{definition}

We use the simple notation $\sigma_{\mu\nu}$ for the $d\times d$ matrix with elements the $\sigma_{\mu\nu}(\hat s,\hat r)$
(i.e. we omit the indices $\hat s,\hat r=0,...,d-1$).
As explained in a general context earlier, the diagonal  $\sigma_{\mu\mu}$ are density matrices for the orbits of our coherent states (and are used later in section 6).
They obey the relations
\begin{eqnarray}\label{VV2}
\frac{d^2}{n}\sum _{\mu=0}^{\mathfrak N}\sigma_{\mu\mu}={\bf 1}_d;\;\;\;\sum _{\mu\ne \nu}\sigma_{\mu\nu}=\sum _{\mu> \nu}(\sigma_{\mu\nu}+\sigma_{\mu \nu}^\dagger) =0.
\end{eqnarray}
The first relation is equivalent to the resolution of the identity in Eq.(\ref{1}).
The second relation will be verified in all examples later.

\begin{definition}
The matrix of overlaps of the coherent states in the $\mu$-orbit with the coherent states in the $\nu$-orbit, is the $d\times d$ circulant matrix
\begin{eqnarray}\label{ABC1}
{\cal T}_{\mu \nu}(\hat r, \hat s)=\langle a_z(\hat r,\mu)\ket{a_z(\hat s, \nu)}=\langle a_z(0,\mu)|X^{\hat s-\hat r}\ket{a_z(0, \nu)};\;\;\;\mu, \nu=0,...,{\mathfrak N};\;\;\;\hat r, \hat s\in{\mathbb Z}_d
\end{eqnarray}
\end{definition}
It is easily seen that ${\cal T}_{\mu \nu}(\widehat {r+k}, \widehat {s+k})={\cal T}_{\mu \nu}(\hat r, \hat s)$ and therefore this matrix is circulant.
The following property is perhaps unusual, and relies on the fact that the orbits are related to the cyclic group ${\cal G}_d$.
\begin{proposition}\label{pro55}
The matrix of overlaps of the coherent states ${\cal T}_{\mu \nu}(\hat r, \hat s)$, and 
their orbit matrices  $\sigma_{\mu\nu}(\hat r, \hat s)$ (times $d$), are both $d\times d$ circulant matrices transpose  to each other with respect to $\hat r, \hat s$ (denoted with $T$):
\begin{eqnarray}\label{EQ}
{\cal T}_{\mu \nu}(\hat r, \hat s)=d{\sigma}^T_{\mu \nu}(\hat r, \hat s);\;\;\;{\sigma}^T_{\mu \nu}(\hat r, \hat s)={\sigma}_{\mu \nu}(\hat s, \hat r)
\end{eqnarray}

\end{proposition}
\begin{proof}
We express the right hand side as
\begin{eqnarray}
d\sigma_{\mu\nu}(\hat s, \hat r)=\sum_{\hat q=0}^{d-1}\bra{a_z(0, \mu)}X^{-\hat q-\hat r}|X;0\rangle\langle X; 0|X^{\hat q+\hat s}\ket{a_z(0,\nu)},
\end{eqnarray}
and we prove that
\begin{eqnarray}
\sum_{\hat q=0}^{d-1}X^{-\hat q-\hat r}|X;0\rangle\langle X; 0|X^{\hat q+\hat s}=X^{\hat s-\hat r}.
\end{eqnarray}
This reduces to 
\begin{eqnarray}
\sum_{\hat q=0}^{d-1}|X;\hat q+\hat r\rangle\langle X; \hat q+\hat s|=X^{\hat s-\hat r},
\end{eqnarray}
which is easily seen to be true.
This proves the equality in Eq.(\ref{EQ}).

\end{proof}
For $\mu\ne\nu$, the diagonal elements of these matrices are equal to zero (Eq.(\ref{Y6})):
\begin{eqnarray}\label{E100}
{\cal T}_{\mu \nu}(\hat r, \hat r)=d\sigma_{\mu\nu}(\hat r, \hat r)=\delta_{\mu\nu}.
\end{eqnarray}

\item
{\bf Discrete isotropy:}
The set of $n$ probabilities
\begin{eqnarray}
A_r=\{|\bra{a_z(r)}a_z(s)\rangle|^2\;|\;s=0,...n-1\},
\end{eqnarray}
is the same for all $r$.
Therefore the following sum does not depend on $r$:
 \begin{eqnarray}\label{73}
S(\nu)=\sum_{s=0}^{n-1}|\bra{a_z(r)}a_z(s)\rangle|^\nu;\;\;\;\nu=1,2,3,....
\end{eqnarray}

\item{\bf The projector of  overlaps of the coherent states is  `reproducing kernel':}
The overlaps between coherent states (times $\frac{d}{n}$) is a $n\times n$ projector (which is also written as a block matrix):
\begin{eqnarray}\label{proj}
\Pi_{rs}=\frac{d}{n}\langle a_z(r)\ket{a_z(s)};\;\;\;\Pi=\frac{d}{n}\begin{pmatrix}
{\cal T}_{00}&\cdots&{\cal T}_{0{\mathfrak N}}\\
\vdots&\vdots&\vdots\\
{\cal T}_{{\mathfrak N}0}&\cdots&{\cal T}_{{\mathfrak N}{\mathfrak N}}\\
\end{pmatrix}=\frac{d^2}{n}\begin{pmatrix}
{\sigma}_{00}^T&\cdots&{\sigma}_{0{\mathfrak N}}^T\\
\vdots&\vdots&\vdots\\
{\sigma}_{{\mathfrak N}0}^T&\cdots&{\sigma}_{{\mathfrak N}{\mathfrak N}}^T\\
\end{pmatrix}
;\;\;\;
{\rm rank}(\Pi)=d.
\end{eqnarray}
Here we use the simple notation ${\cal T}_{\mu\nu}$ for the $d\times d$ matrix with elements the ${\cal T}_{\mu\nu}(\hat s,\hat r)$
(i.e. we omit the indices $\hat s,\hat r=0,...,d-1$).
We prove that $\Pi^2=\Pi$ using the resolution of the identity:
\begin{eqnarray}
\sum _{s=0}^{n-1}\Pi_{rs}\Pi_{st}=\left(\frac{d}{n}\right)^2\sum _{s=0}^{n-1}\langle a_z(r)\ket{a_z(s)}\langle a_z(s)\ket{a_z(t)}=\frac{d}{n}\langle a_z(r)\ket{a_z(t)}=\Pi_{rt}.
\end{eqnarray}
$\Pi$ has the eigenvalues $1$ (with multiplicity $d$) and $0$ (with multiplicity $n-d$).

The discrete isotropy property implies that the
\begin{eqnarray}
\sum _{s=0}^{n-1}|\Pi _{rs}|^\nu=\left(\frac{d}{n}\right )^\nu S(\nu)=\left(\frac{d}{n}\right )^\nu \sum_{s=0}^{n-1}|\bra{a_z(r)}a_z(s)\rangle|^\nu;\;\;\;\nu=1,2,3,....
\end{eqnarray}
does not depend on $r$. 
Also Eq.(\ref{OR}) implies that 
\begin{eqnarray}\label{ORR}
\Pi_{r,r+d\mu}=\frac{d}{n}\delta_{0\mu};\;\;\;\mu=0,...,{\mathfrak N}
\end{eqnarray}
Therefore we have ${\mathfrak N}=\frac{n}{d}-1$ zeros in each row of the projector $\Pi$.

A $n$-tuple representation of states in $H(d)$ analogous to the Bargmann representation is discussed in section \ref{sec41}.
The role of the projector $\Pi$ as a `reproducing kernel' is discussed later (Eq.(\ref{49})).

\item
{\bf  The coherent states are columns of a semi-unitary matrix:}
Let $M$ be a $d\times n$ matrix that has the $d$-dimensional vectors $\ket{a_z(r)}$ in the position basis as columns, times $\sqrt{\frac{d}{n}}$. Then 
$MM^\dagger={\bf 1}_d$ is the resolution of the identity, and $M^\dagger M=\Pi$:
\begin{eqnarray}
M=\sqrt{\frac{d}{n}}(\langle X;j\ket{a_z(r)});\;\;\;M^\dagger M=\Pi;\;\;\;MM^\dagger={\bf 1}_d.
\end{eqnarray}
We call semi-unitary the matrix $M$, in the sense that $M^\dagger M$ is a projector (which might be viewed as a unit matrix within the subspace that it projects into).

\end{itemize}

Overall, our coherent states are highly symmetric. Apart from invariance under the group ${\cal G}_d$ and the discrete isotropy, 
relations like Eqs.(\ref{VV2}), (\ref{EQ}), indicate a high degree of discrete symmetry.
Below we give three examples where $(d,n)$ takes the values $(3,6)$, $(4,8)$, $(4,12)$.
We use the notation of the present section in all these examples, so that the analogy is clear.

\subsection{First example: the set ${\cal C}(3,6,z)$}\label{sec1}
In $H(3)$ we consider the set ${\cal C}(3,6,z)$ with the coherent states (with $|z|=1$ and expressed in the position basis):
\begin{eqnarray}\label{vec}
&&\ket{a_z(0)}=\ket{a_z(0,0)}=\frac{1}{\sqrt{2}}
\begin{pmatrix}
1\\z\\ 0
\end{pmatrix};\;\;
\ket{a_z(1)}=\ket{a_z(1,0)}=\frac{1}{\sqrt{2}}
\begin{pmatrix}
z\\0\\ 1
\end{pmatrix};\;\;
\ket{a_z(2)}=\ket{a_z(2,0)}=\frac{1}{\sqrt{2}}
\begin{pmatrix}
0 \\1\\ z
\end{pmatrix}\nonumber\\
&&\ket{a_z(3)}=\ket{a_z(0,1)}=\frac{1}{\sqrt{2}}
\begin{pmatrix}
1 \\-z\\ 0
\end{pmatrix};\;\;
\ket{a_z(4)}=\ket{a_z(1,1)}=\frac{1}{\sqrt{2}}
\begin{pmatrix}
-z\\0\\ 1
\end{pmatrix};\;\;
\ket{a_z(5)}=\ket{a_z(2,1)}=\frac{1}{\sqrt{2}}
\begin{pmatrix}
0 \\1\\ -z
\end{pmatrix}.
\end{eqnarray}
The relation between the two notations is  based on $r=\hat r+3\mu$ with $\hat r \in\mathbb {Z}_3$ and $\mu=0,1$ and $r=0,...,5$.
These states obey the resolution of the identity:
\begin{eqnarray}\label{res15}
\frac{1}{2}\sum_{r=0}^5\ket{a_z(r)}\bra{a_z(r)}={\bf 1}_3.
\end{eqnarray}

The set ${\cal C}(3,6,z)$ is invariant under transformations in the group ${\cal G}_3$.
The action of the group ${\cal G}_3$  leads to two orbits:
\begin{eqnarray}\label{O1}
&&\ket{a_z(0)}\xrightarrow{X} \ket{a_z(1)}\xrightarrow{X} \ket{a_z(2)}\xrightarrow{X} \ket{a_z(0)};\nonumber\\
&&\ket{a_z(3)}\xrightarrow{X} \ket{a_z(4)}\xrightarrow{X} \ket{a_z(5)}\xrightarrow{X} \ket{a_z(3)},
\end{eqnarray}
The set ${\cal C}(3,6,z)$ is partitioned into the following equivalence classes that contain states in the same orbit (we use here the `pair of indices notation'):
\begin{eqnarray}
{\cal C}_\mu (3,6,z)=\{\ket{a_z(0,\mu)}, \ket{a_z(1,\mu)}, \ket{a_z(2,\mu)}\};\;\;\;\mu=0,1
\end{eqnarray}
The orbit  matrices (Eq.(\ref{ABC})) are the circulant matrices
\begin{eqnarray}\label{35}
&&\sigma_{00}=\frac{1}{3}{\bf 1}_3+\frac{1}{6}(z^*X+zX^\dagger)\nonumber\\
&&\sigma_{11}=\frac{1}{3}{\bf 1}_3-\frac{1}{6}(z^*X+zX^\dagger)\nonumber\\
&&\sigma_{01}=\sigma_{10}^\dagger=\frac{1}{6}(z^*X-zX^\dagger)\nonumber\\
&&\frac{3}{2}(\sigma_{00}+\sigma_{11})={\bf 1}_3;\;\;\;\sigma_{01}+\sigma_{10}=0;\;\;\;{\rm Tr}(\sigma_{\mu\nu})=\delta_{\mu\nu}.
\end{eqnarray}
We easily verify Eqs.(\ref{AAD}), (\ref{VV2}), (\ref{E100}). 
The diagonal ones $\sigma_{\mu\mu}$ are density matrices describing the two orbits, and are important for section 6.

In this example the matrix $M$ is
\begin{eqnarray}\label{AQ1}
M=\frac{1}{2}
\setcounter{MaxMatrixCols}{12}
\begin{pmatrix}
1&z&0&1&-z&0\\
z&0&1&-z&0&1\\
0&1&z&0&1&-z
\end{pmatrix};\;\;\;MM^\dagger={\bf 1}_3;\;\;\;|z|=1.
\end{eqnarray}
and the projector of the overlaps of the coherent states is
\begin{eqnarray}\label{P1}
\Pi=\left(\frac{1}{2}\langle a_z(r)\ket{a_z(s)}\right)=M^\dagger M=\frac{1}{4}
\begin{pmatrix}
2&z&z^*&0&-z&z^*\\
z^*&2&z&z^*&0&-z\\
z&z^*&2&-z&z^*&0\\
0&z&-z^*&2&-z&-z^*\\
-z^*&0&z&-z^*&2&-z\\
z&-z^*&0&-z&-z^*&2
\end{pmatrix}.
\end{eqnarray}
Here ${\rm rank}(\Pi)=3$.
$\Pi$ has the eigenvalues $1$ (with multiplicity $3$) and $0$ (with multiplicity $3$).
We easily confirm the validity of Eq.(\ref{ORR}) in this example.

The $|\Pi_{rs}|=\frac{1}{2}|\langle a_z(r)\ket{a_z(s)}|$ in any of the rows takes the values  $\frac{1}{2}, \frac{1}{4}, 0$ with multiplicities $1, 4, 1$, correspondingly.
This confirms the discrete isotropy property and it can be used to prove that in this example
 \begin{eqnarray}
S(\nu)=\sum_{s=0}^5|\bra{a_z(r)}a_z(s)\rangle|^\nu=2^\nu\sum_{s=0}^5 |\Pi_{rs}|^\nu=1+\frac{1}{2^{\nu-2}};\;\;\;\nu=1,2,3,....
\end{eqnarray}
$S(\nu)$ does not depend on the index $r$.

$\Pi$ can be written as a block matrix as:
\begin{eqnarray}
\Pi=\frac{3}{2}
\begin{pmatrix}
\sigma^T_{00}&\sigma^T_{01}\\
\sigma^T_{10}&\sigma^T_{11}
\end{pmatrix}=\frac{1}{2}
\begin{pmatrix}
{\cal T}_{00}&{\cal T}_{01}\\
{\cal T}_{10}&{\cal T}_{11}
\end{pmatrix}
\end{eqnarray}
This result  expresses the projector in terms of two parts.
The first part is related to the coherent states in the first orbit and the second part is related to the coherent states in the second orbit.
There are also cross terms.

\subsection{Second example: the set ${\cal C}(4,8,z)$}\label{sec2}
In this example the $4 \times 8$ matrix $M$ is
\begin{eqnarray}\label{AQ2}
M=\frac{1}{2}
\setcounter{MaxMatrixCols}{12}
\begin{pmatrix}
z&1&0&0&z&-1&0&0\\
1&0&0&z&-1&0&0&z\\
0&0&z&1&0&0&z&-1\\
0&z&1&0&0&z&-1&0
\end{pmatrix};\;\;\;|z|=1;\;\;\;MM^\dagger={\bf 1}_4.
\end{eqnarray}
The states $\ket{a_z(0)},...,\ket{a_z(7)}$ (in the position basis) are the columns of this matrix multiplied by $\sqrt{2}$:
\begin{eqnarray}
\ket{a_z(0)}=\ket{a_z(0,0)}=\frac{1}{\sqrt{2}}
\begin{pmatrix}
z\\1\\0\\0
\end{pmatrix};\cdots;
\ket{a_z(7)}=\ket{a_z(3,1)}=\frac{1}{\sqrt{2}}
\begin{pmatrix}
0\\z\\-1\\0
\end{pmatrix}.
\end{eqnarray}

They obey the resolution of the identity:
\begin{eqnarray}\label{res15}
\frac{1}{2}\sum_{r=0}^7\ket{a_z(r)}\bra{a_z(r)}={\bf 1}_4.
\end{eqnarray}

The set ${\cal C}(4,8,z)$ of these states, is invariant under transformations in the group ${\cal G}_4$.
The action of the group ${\cal G}_4$ on the set ${\cal C}(4,8,z)$ leads to two orbits:
\begin{eqnarray}\label{O2}
&&\ket{a_z(0)}\xrightarrow{X} \ket{a_z(1)}\xrightarrow{X} \ket{a_z(2)}\xrightarrow{X} \ket{a_z(3)}\xrightarrow{X} \ket{a_z(0)};\nonumber\\
&&\ket{a_z(4)}\xrightarrow{X} \ket{a_z(5)}\xrightarrow{X} \ket{a_z(6)}\xrightarrow{X} \ket{a_z(7)}\xrightarrow{X} \ket{a_z(4)}.
\end{eqnarray}
Therefore it partitions ${\cal C}(4,8,z)$ into the following equivalence classes that contain states in the same orbit (we use here the `pair of indices notation'):
\begin{eqnarray}
{\cal C}_\mu(4,8,z)=\{\ket{a_z(0,\mu)}, \ket{a_z(1,\mu)}, \ket{a_z(2,\mu)}, \ket{a_z(3,\mu)}\};\;\;\;\mu=0,1.
\end{eqnarray}
The orbit matrices are the circulant matrices
\begin{eqnarray}\label{45}
&&\sigma_{00}=\frac{1}{4}{\bf 1}_4+\frac{1}{8}(zX+z^*X^\dagger);\;\;\;
\sigma_{11}=\frac{1}{4}{\bf 1}_4-\frac{1}{8}(zX+z^*X^\dagger)\nonumber\\
&&\sigma_{01}=\sigma_{10}^\dagger=\frac{1}{8}(zX-z^*X^\dagger)\nonumber\\
&&2(\sigma_{00}+\sigma_{11})={\bf 1}_4;;\;\;\;\sigma_{01}+\sigma_{10}=0\;\;\;{\rm Tr}(\sigma_{\mu\nu})=\delta_{\mu\nu}.
\end{eqnarray}
We easily verify Eqs.(\ref{AAD}), (\ref{VV2}), (\ref{E100}). 
The diagonal ones $\sigma_{\mu\mu}$ are density matrices describing the two orbits, and are used in section 6.

In this example  the projector $\Pi=M^\dagger M$ of the overlaps of the coherent states, is:
\begin{eqnarray}\label{P2}
\Pi=\left(\frac{1}{2}\langle a_z(r)\ket{a_z(s)}\right)=M^\dagger M=\frac{1}{4}
\begin{pmatrix}
2&z^*&0&z&0&-z^*&0&z\\
z&2&z^*&0&z&0&-z^*&0\\
0&z&2&z^*&0&z&0&-z^*\\
z^*&0&z&2&-z^*&0&z&0\\
0&z^*&0&-z&2&-z^*&0&-z\\
-z&0&z^*&0&-z&2&-z^*&0\\
0&-z&0&z^*&0&-z&2&-z^*\\
z^*&0&-z&0&-z^*&0&-z&2
\end{pmatrix}
\end{eqnarray}$
\Pi$ has the eigenvalues $1$ (with multiplicity $4$) and $0$ (with multiplicity $4$).
We easily confirm the validity of Eq.(\ref{ORR}) in this example.

The $|\Pi_{rs}|=\frac{1}{2}|\langle a_z(r)\ket{a_z(s)}|$ in any of the rows takes the values  $\frac{1}{2}, \frac{1}{4}, 0$ with multiplicities $1, 4, 3$, correspondingly.
This confirms the discrete isotropy property and it can be used to prove that in this example
 \begin{eqnarray}
S(\nu)=\sum_{s=0}^7|\bra{a_z(r)}a_z(s)\rangle|^\nu=2^\nu\sum_{s=0}^7 |\Pi_{rs}|^\nu=1+\frac{1}{2^{\nu-2}};\;\;\;r=1,2,3,....
\end{eqnarray}
$S(\nu)$ does not depend on the index $r$.

$\Pi$ can be written as a block matrix in terms of the orbit operators:
\begin{eqnarray}
\Pi=2
\begin{pmatrix}
\sigma^T_{00}&\sigma^T_{01}\\
\sigma^T_{10}&\sigma^T_{11}
\end{pmatrix}=\frac{1}{2}
\begin{pmatrix}
{\cal T}_{00}&{\cal T}_{01}\\
{\cal T}_{10}&{\cal T}_{11}
\end{pmatrix}
\end{eqnarray}

\subsection{Third example: the set ${\cal C}(4,12,z)$}\label{sec3}
In this example the $4 \times 12$ matrix $M$ is
\begin{eqnarray}\label{AQ3}
M=\frac{1}{3}
\setcounter{MaxMatrixCols}{12}
\begin{pmatrix}
z&1&1&0&z&\omega&\omega^2&0&z&\omega^2&\omega&0\\
1&1&0&z&\omega&\omega^2&0&z&\omega^2&\omega&0&z\\
1&0&z&1&\omega^2&0&z&\omega&\omega&0&z&\omega^2\\
0&z&1&1&0&z&\omega&\omega^2&0&z&\omega^2&\omega
\end{pmatrix};\;\;|z|=1;\;\;\omega=\exp\left(\frac{2\pi i}{3}\right );\;\;MM^\dagger={\bf 1}_4.
\end{eqnarray}
The states $\ket{a_z(0)},...,\ket{a_z(11)}$ (in the position basis) are the columns of this matrix multiplied by $\sqrt{3}$:
\begin{eqnarray}
\ket{a_z(0)}=\ket{a_z(0,0)}=\frac{1}{\sqrt{3}}
\begin{pmatrix}
z\\1\\1\\0
\end{pmatrix};\cdots;
\ket{a_z(11)}=\ket{a_z(3,2)}=\frac{1}{\sqrt{3}}
\begin{pmatrix}
0\\z\\\omega^2\\\omega
\end{pmatrix}.
\end{eqnarray}
They obey  the resolution of the identity:
\begin{eqnarray}\label{res15}
\frac{1}{3}\sum_{r=0}^{11}\ket{a_z(r)}\bra{a_z(r)}={\bf 1}_4.
\end{eqnarray}

The set ${\cal C}(4,12,z)$ of these states, is invariant under transformations in the group ${\cal G}_4$.
The action of the group ${\cal G}_4$ on the set ${\cal C}(4,12,z)$ leads to three orbits with the states in the following equivalence classes:
\begin{eqnarray}
{\cal C}_\mu(4,12,z)=\{\ket{a_z(0,\mu)}, \ket{a_z(1,\mu)}, \ket{a_z(2,\mu)}, \ket{a_z(3,\mu)}\};\;\;\;\mu=0,1,2.
\end{eqnarray}
The orbit matrices are the circulant matrices
\begin{eqnarray}\label{GG3}
&&\sigma_{00}=\frac{1}{4}{\bf 1}_4+\frac{1}{12}(z^*+1)X^3+\frac{1}{12}(z+z^*)X^2+\frac{1}{12}(z+1)X\nonumber\\
&&\sigma_{11}=\frac{1}{4}{\bf 1}_4+\frac{1}{12}(z^*+1)\omega X^3+\frac{1}{12}(z\omega+z^*\omega^2)X^2+\frac{1}{12}(z+1)\omega^2X\nonumber\\
&&\sigma_{22}=\frac{1}{4}{\bf 1}_4+\frac{1}{12}(z^*+1)\omega^2 X^3+\frac{1}{12}(z\omega^2+z^*\omega)X^2+\frac{1}{12}(z+1)\omega X\nonumber\\
&&\frac{4}{3}(\sigma_{00}+\sigma_{11}+\sigma_{22})={\bf 1}_4,
\end{eqnarray}
They are needed in section 6. Also
\begin{eqnarray}\label{GG4}
&&12\sigma_{01}=12\sigma_{10}^\dagger=(z^*\omega+\omega^2)X^3+(z^*\omega^2+z)X^2+(z+\omega)X\nonumber\\
&&12\sigma_{02}=12\sigma_{20}^\dagger=(z^*\omega^2+\omega)X^3+(z^*\omega+z)X^2+(z+\omega^2)X\nonumber\\
&&12\sigma_{12}=12\sigma_{21}^\dagger=(z^*\omega^2+1)X^3+(z^*\omega+z\omega)X^2+(z\omega^2+1)X\nonumber\\
&&{\rm Tr}(\sigma_{\mu\nu})=\delta_{\mu\nu}.
\end{eqnarray}
We easily verify Eqs.(\ref{AAD}), (\ref{VV2}), (\ref{E100}).

In this example $\Pi=M^\dagger M=\frac{1}{3}\langle a_z(r)\ket{a_z(s)}$ is a $12\times 12$ projector. The elements of $9\Pi_{rs}$  are given in table \ref{t1}.
$\Pi$ has the eigenvalues $1$ (with multiplicity $4$) and $0$ (with multiplicity $8$).
We easily confirm the validity of Eq.(\ref{ORR}) in this example.

The $|\Pi_{rs}|=\frac{1}{2}|\langle a_z(r)\ket{a_z(s)}|$ in any of the rows takes the values  $\frac{3}{9}, \frac{2}{9}, \frac{1}{9}, 0$ with multiplicities $1, 3, 6, 2$, correspondingly.
This confirms the discrete isotropy property and it can be used to prove that in this example
 \begin{eqnarray}
S(\nu)=\sum_{s=0}^{11}|\bra{a_z(r)}a_z(s)\rangle|^\nu=3^\nu\sum_{s=0}^{11} |\Pi_{rs}|^\nu=1+\frac{2^\nu+2}{3^{\nu-1}};\;\;\;r=1,2,3,....
\end{eqnarray}
$S(\nu)$ does not depend on the index $r$.

$\Pi$ can be written as a block matrix as
\begin{eqnarray}
\Pi=\frac{4}{3}
\begin{pmatrix}
\sigma^T_{00}&\sigma^T_{01}\sigma^T_{02}\\
\sigma^T_{10}&\sigma^T_{11}\sigma^T_{12}\\
\sigma^T_{20}&\sigma^T_{21}\sigma^T_{00}
\end{pmatrix}=\frac{1}{3}
\begin{pmatrix}
{\cal T}_{00}&{\cal T}_{01}&{\cal T}_{02}\\
{\cal T}_{10}&{\cal T}_{11}&{\cal T}_{12}\\
{\cal T}_{20}&{\cal T}_{21}&{\cal T}_{00}
\end{pmatrix}
\end{eqnarray}

\section{$n$-tuple representation of states in $H(d)$}\label{sec41}
The formalism of this section applies to all sets of coherent states ${\cal C}(d,n,z)$  discussed earlier.
We consider an arbitrary state in $H(d)$, in the position basis:
\begin{eqnarray}
\ket{f}=\begin{pmatrix}
{f}_0\\
\vdots\\
{f}_{d-1}
\end{pmatrix};\;\;\;\sum_{i=0}^{d-1}| {f}_i|^2=1,
\end{eqnarray}
written here in the `usual' $d$-tuple representation.
Using the resolution of the identity in Eq.(\ref{1}) we expand it as
\begin{eqnarray}\label{60}
\ket{f}=\sqrt{\frac{d}{n}}\sum_{r=0}^{n-1} {\widetilde f}(r) \ket{a_z(r)};\;\;\;{\widetilde f}(r)=\sqrt{\frac{d}{n}}\bra {a_z(r)} f \rangle=\sum _{j=0}^{d-1}(M^\dagger)_{rj}f_j\;\;\;\sum_{r=0}^{n-1}|{\widetilde f}(r)|^2=1.
\end{eqnarray}
The tilde in the notation indicates $n$-tuple representation. 
The last equation is proved using the resolution of the identity.
Each $|\bra {a_z(r)} f \rangle|^2$ is a probability, but they are not independent (because the coherent states are not orthogonal to each other) and for this reason their sum is $\frac{n}{d}$.
In the `pair of indices notation'  the above expansion is written as
\begin{eqnarray}\label{59}
\ket{f}=\sqrt{\frac{d}{n}}\sum_{\mu=0}^{\frak{N}}\sum_{\hat r=0}^{d-1} {\widetilde f}(\hat r, \mu)\ket{a_z(\hat r, \mu)};\;\;\;{\widetilde f}(\hat r, \mu)=\sqrt{\frac{d}{n}}\bra {a_z(\hat r, \mu)} f \rangle;\;\;\;
\sum_{\mu=0}^{\frak{N}}\sum_{\hat r=0}^{d-1} |{\widetilde f}(\hat r, \mu)|^2=1
\end{eqnarray}

The `$n$-tuple representation' is the analogue of the Bargmann representation of the harmonic oscillator. They are both based on states that resolve the identity and are not orthogonal to each other.
Using the resolution of the identity we prove the `reproducing kernel' relation
\begin{eqnarray}\label{49}
\sum_{r=0}^{n-1} \Pi_{sr}{\widetilde f}(r)= {\widetilde f}(s).
\end{eqnarray}
The 
\begin{eqnarray}
\begin{pmatrix}
{f}_0\\
\vdots\\
{f}_{d-1}
\end{pmatrix}\;\rightarrow
\begin{pmatrix}
{\widetilde f}(0)\\
\vdots\\
{\widetilde f}(n-1)
\end{pmatrix};\;\;\;\sum_{i=0}^{d-1}|f_i|^2=\sum_{r=0}^{n-1}|{\widetilde f}(r)|^2=1.
\end{eqnarray}
can be viewed as a  map from $H(d)$ into a subspace of $H(n)$ (which is isomorphic to $H(d)$)
and preserves the scalar product
\begin{eqnarray}
\langle g\ket{f}=\sum _{i=1}^dg_i^*f_i=\sum _{i=1}^d ({\widetilde g}_i)^*{\widetilde f}_i.
\end{eqnarray}
The merit of working in a larger space ($n>d$)  is redundancy.
Noisy coefficients ${\widetilde f}(r)$ will reproduce the state more accurately, than noisy coefficients $f_i$.  
This is similar to ordinary language where due to redundancy, a small spelling mistake does not change the meaning
(as quantified by Shannon and later by many others).
We do not explore this direction in the present paper.

A density matrix $\rho$ will be represented with the $n\times n$ matrix
\begin{eqnarray}
&&\rho=\frac{d}{n}\sum_{r,s}{\widetilde F}(r,s) \ket{a_z(r)}\bra{a_z(s)};\;\;\;{\widetilde F}(r,s)=\frac{d}{n}\bra {a_z(r)} \rho\ket{a_z(s)};\;\;\;r,s=0,...,n-1\nonumber\\
&&\sum_{r=0}^{n-1}|{\widetilde F}(r,r)|^2=1.
\end{eqnarray}
In the pair of indices notation this is written as
\begin{eqnarray}
&&\rho=\frac{d}{n}\sum_{r,s}{\widetilde F}(\widehat r, \mu; \widehat s, \nu)\ket{a_z(\widehat r, \mu)}\bra{a_z(\widehat s, \nu)};\;\;\;{\widetilde F}(\widehat r, \mu; \widehat s, \nu)=\frac{d}{n}\bra {a_z(\widehat r, \mu)} \rho\ket{a_z(\widehat s, \nu)}\nonumber\\
&&\sum_{r=0}^{d-1}\sum_{\mu=0}^{\mathfrak N}|{\widetilde F}(\widehat r, \mu; \widehat r, \mu)|^2=1.
\end{eqnarray}

The eigenvectors of the projector $\Pi$ corresponding to the eigenvalue $1$ span the space $H(d)$, and the 
eigenvectors corresponding to the eigenvalue $0$ span the null space $H(n-d)_{\rm null}$.
Let
\begin{eqnarray}
H(n)=H(d)\oplus H(n-d)_{\rm null}.
\end{eqnarray}
Then for the $n$-tuples in $H(n-d)_{\rm null}$
\begin{eqnarray}
\Pi\ket {g}=0;\;\;\;\ket{g}\in H(n-d)_{\rm null}.
\end{eqnarray}
For the $n$-tuples in $H(d)$ we have the reproducing kernel relation in Eq.(\ref{49}). 
\begin{lemma}\label{L}
We consider the $n$-tuple representation of states $\ket{f}\in H(d)$, with respect to the coherent states in any of the three sets ${\cal C}(d,n,z)$ described above. For generic values of 
the parameter $z=\exp(i\theta)$, we
cannot have all $|{\widetilde f}(r)|^2=\frac{1}{n}$.
\end{lemma}
\begin{proof}
\mbox{}
\begin{itemize}
\item{\bf The set ${\cal C}(3,6,z)$:}
Eq.(\ref{60}) with the matrix $M$ in Eq.(\ref{AQ1}) gives the following $6$ relations (with $|z|=1$):
\begin{eqnarray}\label{DD1}
&&|f_0+z^*f_1|=2|{\widetilde f}(0)|;\;\;\;|z^*f_0+f_2|=2|{\widetilde f}(1)|;\;\;\;|f_1+z^*f_2|=2|{\widetilde f}(2)|\nonumber\\
&&|f_0-z^*f_1|=2|{\widetilde f}(3)|;\;\;\;|-z^*f_0+f_2|=2|{\widetilde f}(4)|;\;\;\;|f_1-z^*f_2|=2|{\widetilde f}(5)|.
\end{eqnarray}
If all $|{\widetilde f}(r)|^2=\frac{1}{6}$ it follows that all $|f_i|^2=\frac{1}{3}$, but this cannot satisfy all relations in Eq.(\ref{DD1}).
Indeed, we need $\Re(f_0zf_1^*)=\Re(z^*f_0f_2^*)=\Re(f_1zf_2^*)=0$ and if $f_i=\frac{1}{\sqrt {3}}\exp(i\phi_i)$ and $z=\exp(i\theta)$ then we need
\begin{eqnarray}
\cos (\phi_0-\phi_1+\theta)=\cos (\phi_1-\phi_2+\theta)=\cos (\phi_2-\phi_0+\theta)=0.
\end{eqnarray}
This is not possible for generic $\theta$. It is only possible for special values of $\theta$ (e.g., $\theta=\frac{\pi}{2}$ and $\phi_0=\phi_1=\phi_2=\phi_3$).

\item{\bf The set ${\cal C}(4,8,z)$:}
Eq.(\ref{60}) with the matrix $M$ in Eq.(\ref{AQ2}) gives $8$ relations (with $|z|=1$):
\begin{eqnarray}\label{DD2}
&&|f_0+z^*f_1|=2|{\widetilde f}(0)|;\;\;\;|f_0-z^*f_1|=2|{\widetilde f}(4)|;\;\;\;{\rm etc}.
\end{eqnarray}
If all $|{\widetilde f}(r)|^2=\frac{1}{8}$ it follows that $|f_0|^2+|f_1|^2=\frac{1}{2}$, and through symmetry we prove that all $|f_i|^2=\frac{1}{4}$.
We also have the relations
\begin{eqnarray}
\Re(z^*f_0f_1^*)=\Re(f_0zf_3^*)=\Re(z^*f_2f_3^*)=\Re(z^*f_1f_2^*)=0.
\end{eqnarray}
If $f_i=\frac{1}{2}\exp(i\phi_i)$ and $z=\exp(i\theta)$ then we need
\begin{eqnarray}
\cos (\phi_0-\phi_1-\theta)=\cos (\phi_1-\phi_2-\theta)=\cos (\phi_2-\phi_3-\theta)=\cos (\phi_3-\phi_0-\theta)=0.
\end{eqnarray}
This is not possible for generic $\theta$. It is only possible 
for special values of $\theta$ (e.g., $\theta=\frac{\pi}{2}$ and $\phi_0=\phi_1=\phi_2=\phi_3$).

\item{\bf The set ${\cal C}(4,12,z)$:}
Eq.(\ref{60}) with the matrix $M$ in Eq.(\ref{AQ3}) gives $12$ relations (with $|z|=1$):
\begin{eqnarray}\label{DD3}
|z^* f_0+f_1+f_2|=3|{\widetilde f}(0)|;\;\;\;|z^* f_0+\omega^2 f_1+\omega f_2|=3|{\widetilde f}(4)|;\;\;\;|z^* f_0+\omega f_1+\omega^2 f_2|=3|{\widetilde f}(8)|;\;\;\;{\rm etc.}
\end{eqnarray}
If all $|{\widetilde f}(r)|^2=\frac{1}{12}$ it  follows that 
$|f_0|^2+|f_1|^2+|f_2|^2=\frac{3}{4}$, and through symmetry we prove that all $|f_i|^2=\frac{1}{4}$.
We also get the $12$ equations
\begin{eqnarray}
&&z^*f_0f_1^*+f_1f_2^*+f_2zf_0^*+zf_0^*f_1+f_1^*f_2+f_2^*z^*f_0=0\nonumber\\
&&z^*f_0f_1^*\omega+zf_0^*f_1\omega^2+f_2zf_0^*\omega+z^*f_0f_2^*\omega^2+f_1f_2^*\omega+f_2f_1^*\omega^2=0\nonumber\\
&&z^*f_0f_1^*\omega^2+zf_0^*f_1\omega +f_2zf_0^*\omega^2+z^*f_0f_2^*\omega+f_1^*f_2\omega+f_2^*f_1\omega^2=0;\;\;\;{\rm e.t.c.}
\end{eqnarray}
We multiply the first three equations by $1, \omega^2, \omega$ correspondingly, and adding them we get:
\begin{eqnarray}\label{111}
z^*f_0f_1^*+f_1f_2^*+f_2zf_0^*=0.
\end{eqnarray}
In a similar way we also get
\begin{eqnarray}
f_0f_1^*+zf_1f_3^*+z^*f_3f_0^*=0;\;\;\;
f_0f_3^*+zf_3f_2^*+z^*f_2f_0^*=0;\;\;\;
z^*f_1f_2^*+f_2f_3^*+f_3zf_1^*=0.
\end{eqnarray}
These four equations are satisfied only if the phases of the three terms in each equation, differ from each other by $2\pi/3$.
If $f_i=\frac{1}{2}\exp(i\phi_i)$  and $z=\exp(i\theta)$, the three phases in the terms of these equations are
\begin{eqnarray}
(\phi_0-\phi_1-\theta, \phi_1-\phi_2, \phi_2-\phi_0+\theta);\;\;\;(\phi_0-\phi_1, \phi_1-\phi_3+\theta, \phi_3-\phi_0-\theta)\nonumber\\
(\phi_0-\phi_3, \phi_3-\phi_2+\theta, \phi_2-\phi_0-\theta);\;\;\;(\phi_1-\phi_2-\theta, \phi_2-\phi_3, \phi_3-\phi_1+\theta).
\end{eqnarray}
We can easily see that for a generic value of $\theta$ these phases cannot differ from each other by $2\pi/3$.
But for $\theta=0$ (and $\phi_0=\phi_2=0$ and $\phi_1=\phi_3=\frac{2\pi}{3}$) or $\theta=\pm \frac{2\pi}{3}$ (and $\phi_0=\phi_1=\phi_2=\phi_3$) this is possible.

\end{itemize}
\end{proof}

\subsection{Stroboscopic time evolution: expectation values of the orbit density matrices}

We next consider stroboscopic time evolution with the operator $X^t$ where $t$ is an integer. In the `pair of indices notation' we get
\begin{eqnarray}
X^t\ket{f}=\sqrt{\frac{d}{n}}\sum_{\mu=0}^{\frac{n}{d}-1}\sum_{\hat r=0}^{d-1} {\widetilde f}(\hat r, \mu)X^t \ket{a_z(\hat r, \mu)}=\sqrt{\frac{d}{n}}\sum_{\mu=0}^{\frac{n}{d}-1}\sum_{\hat r=0}^{d-1} {\widetilde f}(\hat r,\mu) \ket{a_z(\widehat {r+t}, \mu)}
\end{eqnarray}
It is seen that at any discrete time $t$, the set of $d$ coefficients ${\cal T}_\mu(\ket{f})=\{{\widetilde f}(\hat r,\mu)\}$  (with fixed $\mu$ and $\hat r=0,...,d-1$) associated to coherent states in the orbit ${\cal C}_\mu(d,n)$ is invariant.
Related to this is the fact that the expectation values of the orbit density matrices $\sigma_\mu$, are invariant under stroboscopic (discrete) time evolution:
\begin{eqnarray}
\bra{f}(X^\dagger)^t\sigma_\mu X^t\ket{f}=\bra{f}\sigma_\mu \ket{f}=\frac{n}{d^2}\sum_{\hat r=0}^{d-1} |{\widetilde f}(\hat r,\mu)|^2;\;\;\;\sum_{\mu}\bra{f}\sigma_\mu \ket{f}=\frac{n}{d^2};\;\;\;0\le \bra{f}\sigma_\mu \ket{f}\le 1.
\end{eqnarray}
The first equality is based on the fact that $X$ commutes with the orbit operators.
The $\bra{f}\sigma_\mu \ket{f}$ takes values in $[0,1]$ and indicates how close  the state $\ket{f}$ is to the states in the orbit ${\cal C}_\mu(d,n)$.

\begin{example}
In $H(3)$ we consider the coherent states ${\cal C}(3,6,z)$ and 
\begin{eqnarray}
\theta=\ket{f_1}\bra{f_1};\;\;\;\ket{f_1}=\frac{1}{\sqrt{14}}
\begin{pmatrix}
1\\-3\\2
\end{pmatrix};
\end{eqnarray}
Using the orbit operators in Eq.(\ref{35}) we find
\begin{eqnarray}
 {\rm Tr}(\sigma _{00}\theta)=\frac{1}{3}-\frac{z+z^*}{12};\;\;\;{\rm Tr}(\sigma _{11}\theta)=\frac{1}{3}+\frac{z+z^*}{12}
   \end{eqnarray}
The  ${\rm Tr}(\sigma _{00}\theta)$ is the average expectation value of the operator $\theta$ during the stroboscopic time evolution of 
the coherent states in the orbit ${\cal C}_0(3,6,z)$. Similar interpretation can be given for ${\rm Tr}(\sigma _{11}\theta)$.
\end{example}

\section{${\mathfrak Q}$ with the ultra-quantum coherent states takes  values in the Grothendieck region $(1,k_G)$ }\label{sec58}

In this section we present briefly the Grothendieck formalism and express ${\mathfrak Q}$ as the trace of a product of arbitrary (normalised) matrices, because this
 is suitable for quantum mechanical applications.
The present work generalises the examples given in \cite{VOU1} and also gives some new examples (the coherent states in ${\cal C}(4,8,z)$) which lead to 
${\mathfrak Q}$ with  values in the Grothendieck region $(1,k_G)$.
This justifies the term `ultra-quantum' for these coherent states.

Many physical examples use the expression $\bra{e}U\ket{f}$ where $U$ is a unitary operators and 
$\ket{e}, \ket{f}$ are normalised states (e.g., Wigner and Weyl functions).
In example \ref{ex1}, we show that the corresponding ${\mathfrak Q}\le 1$.
In this sense, it is only `rare examples' that give ${\mathfrak Q}$ in the Grothendieck region $(1,k_G)$. 
 
\begin{definition}
$G_n$ is the set of all  $n\times n$ complex matrices $\theta$, such that for all $|a_r|\le 1$, $|b_s|\le 1$
\begin{eqnarray}\label{892}
{\mathfrak C}(\theta)=\left |\sum_{r,s=1}^n\theta _{rs}a_rb_s\right |\le 1
\end{eqnarray}
${\mathfrak C}(\theta)$ is `classical' in the sense that $a_r, b_s$ are scalar quantities.
\end{definition}
If $\Theta$ is an arbitrary $n\times n$ matrix, let
\begin{eqnarray}\label{29}
g(\Theta)=\sup \left \{{\mathfrak C}(\Theta)=\left |\sum_{r,s}\Theta_{rs} a_rb_s\right |;\;\;\;|a_r|\le 1;\;\;\;|b_s|\le 1;\;\;\;r,s=0,...,n-1\right \}.
\end{eqnarray}
Then 
\begin{eqnarray}\label{T6}
\lambda\le \frac{1}{g(\Theta)}\;\rightarrow\;\theta=\lambda \Theta\in G_n.
\end{eqnarray}
It can be proved \cite{VOU1} that $g(\Theta)\le n{\mathfrak s}_{\rm max}$, where  ${\mathfrak s}_{\rm max}$ is the largest singular value of $\Theta$.
For normal matrices ${\mathfrak s}_{\rm max}=e_{\rm max}$, where $e_{\rm max}$ is the is the spectral radius of $\Theta$ (i.e., the largest of all absolute values of its eigenvalues).

We note that:
\begin{itemize}
\item
Given two normalised states $\ket{f}=\sum f_r\ket{X;r}$ and $\ket{e}=\sum e_s\ket{X;s}$, let
\begin{eqnarray}
\Theta_{rs}=\langle X;r\ket{f}\bra{e}X;s\rangle=f_re_s^*.
\end{eqnarray}
Then $g(\Theta)>1$. Indeed 
\begin{eqnarray}
{\mathfrak C}(\Theta)=\left |\sum_{r,s}f_re_s^* a_rb_s\right |;\;\;\;|a_r|\le 1;\;\;\;|b_s|\le 1.
\end{eqnarray}
We choose $a_r, b_s$ such that $f_ra_r=|f_r|$ and $e_s^*b_s=|e_s|$.
Then 
\begin{eqnarray}\label{PO}
g(\Theta)\ge {\mathfrak C}(\Theta)=\sum_r|f_r|\sum _s|e_s|\ge 1.
\end{eqnarray}

\item
$G_n$ is a convex set of $n\times n$ matrices because if $\theta, \phi\in G_n$, then
for $0\le \lambda\le 1$ the matrix $\lambda \theta+ (1-\lambda)\phi$ belongs in $G_n$.

\item
If $\theta \in G_n$ and $U$ is a unitary matrix, then the matrix $U\theta U^\dagger$ might not belong in $G_n$.

\item
If $\theta \in G_n$ then the `enlarged with zeros' matrix
\begin{eqnarray}\label{63}
\widetilde \theta=\begin{pmatrix}
\theta&0_{n,k}\\
0_{k,n}&0_{k,k}
\end{pmatrix}\in G_{n+k}.
\end{eqnarray}
The notation here is self-explanatory.

\end{itemize}

\subsection{Grothendieck theorem}
 We use the `bra-ket' notation for normalised vectors.
The  Grothendieck theorem proves that if Eq.(\ref{892}) holds (i.e., if $\theta \in G_n$), then for all $\lambda _r\ket{u_r},\mu_s\ket{v_s}\in H(n)$ with $\lambda_r, \mu_s\le 1$, the ${\mathfrak Q}(\theta)$ takes values
\begin{eqnarray}\label{344}
{\mathfrak Q}(\theta)=\left |\sum_{r,s}\theta_{rs}\lambda_r\mu_s\bra{u_r}v_s\rangle \right |\le k(n);\;\;\;\theta \in G_n.
\end{eqnarray}
${\mathfrak Q}(\theta)$ is `quantum' in the sense that it replaces the scalar quantities in ${\mathfrak C}(\theta)$ with vectors.
$k(n)$ is the supremum of all ${\mathfrak Q}(\theta)$ (with $\theta\in G_n$), and we refer to it as the Grothendieck bound. It is an increasing function of the dimension $n$:
\begin{eqnarray}\label{65}
n_1\le n_2\;\rightarrow\;k(n_1)\le k(n_2).
\end{eqnarray} 
This is because if $\theta\in G_{n_1}$ and $\ket{u_r}, \ket{v_s}\in H(n_1)$ give the supremum of all ${\mathfrak Q}_1(\theta)$ which is $k(n_1)$,  then the enlarged with zeros matrix $\widetilde \theta\in G_{n_2}$ (Eq.(\ref{63}))
together with the `enlarged with zeros vectors ' 
\begin{eqnarray}
\widetilde {\ket{u_r}}=\begin{pmatrix}
\ket{u_r}\\
0_{n_2-n_1}
\end{pmatrix},
\widetilde {\ket{v_s}}=\begin{pmatrix}
\ket{u_s}\\
0_{n_2-n_1}
\end{pmatrix}
\in H(n_2),
\end{eqnarray} 
will give ${\mathfrak Q}_2(\widetilde \theta)=k(n_1)$.
For more general matrices and vectors in $H(n_2)$ we might get larger values, and therefore follows Eq(\ref{65}).
As $n$ goes to infinity
\begin{eqnarray}
\lim _{n\to \infty}k(n)=k_G,
\end{eqnarray}
where $k_G$ is the complex Grothendieck constant (it is a `global Grothendieck bound'). Its exact value is not known, but it is known\cite{BB1,BB2,BB3} that $1<k_G\le 1.4049$.

The Grothendieck bound $k_G$ is a `mathematical ceiling' of the Hilbert formalism, and in turn of the quantum formalism which is intimately related to it.
The `classical ceiling' for ${\mathfrak C}$ is $1$, while the `quantum ceiling' for  ${\mathfrak Q}$ is $k_G$ (for all dimensions $n$).
In this paper we are mainly interested in  the `Grothendieck region' $(1,k_G)$, which is classically forbidden in the sense that ${\mathfrak C}\le 1$.

\subsection{${\mathfrak Q}$ as the trace of a product of matrices}\label{sec56}

\begin{definition}
${\cal S}_n$ is the set of all $n\times n$ complex matrices $M$ such that ${\cal N}(M)\le 1$, where
\begin{eqnarray}\label{109}
{\cal N}(M)=\max_r \sqrt{\sum _s |M_{rs}|^2}=\max_r \sqrt{(MM^\dagger)_{rr}}.
\end{eqnarray}
\end{definition}
All unitary operators $U$ belong in ${\cal S}_n$ (${\cal N}(U)=1$).

In quantum mechanics we usually express various quantities as the trace of products of matrices.
Ref\cite{VOU1} has expressed  ${\mathfrak Q}$ of Eq.(\ref{344}), as
\begin{eqnarray}\label{77}
{\mathfrak Q}=|{\rm Tr}(\theta VW^\dagger )|;\;\;\;\theta \in G_n;\;\;\;V,W \in {\cal S}_n.
\end{eqnarray}
Here $V$ is a $n\times n$ matrix that has the components of $\mu_s\ket{v_s}$ in the $s$-row, and $W$ is a matrix that has the components of $\lambda_r\ket{u_r}$ in the $r$-row
(therefore $W^\dagger$ has the complex conjugates of the components of $\lambda_r\ket{u_r}$ in the $r$-column).
Consequently $W,V$ are matrices with $d$ row vectors that have norm less or equal to $1$:
\begin{eqnarray}\label{QQ}
\sum_s|W_{rs}|^2\le 1;\;\;\;\sum_s|V_{rs}|^2\le 1.
\end{eqnarray}
Therefore $W,V$ belong to ${\cal S}_n$.

The requirement $\theta \in G_n$ and $V,W \in {\cal S}_n$ in Eq.(\ref{77}) {\bf is not a restriction}, 
because arbitrary matrices with appropriate normalisation will satisfy this requirement.
We have seen this in Eq.(\ref{T6}), and also if $V_1$ is arbitrary $n\times n$ matrix then 
\begin{eqnarray}
\lambda\le \frac{1}{{\cal N}(V_1)}\;\rightarrow\;V=\lambda V_1\in {\cal S}_n.
\end{eqnarray}
We note that if $U$ is a unitary matrix, then in general
\begin{eqnarray}
{\cal N}(UVU^\dagger)\ne {\cal N}(V);\;\;\;g(U\theta U^\dagger)\ne g(\theta).
\end{eqnarray}
Therefore the basis that we use in order to represent an operator as a matrix, is very important.

\begin{example}\label{ex1}
We consider the  $\bra{e}U\ket{f}$ where $U\in {\cal S}_n$ (i.e., ${\cal N}(U)\le 1$) and 
$\ket{e}, \ket{f}$ are normalised states.
A physically important class of operators $U\in {\cal S}_n$ are the unitary operators (in which case ${\cal N}(U)=1$).
We show that the corresponding ${\mathfrak Q}\le 1$.
We choose 
\begin{eqnarray}
\theta=\frac{\ket{f}\bra{e}}{g(\ket{f}\bra{e})}\in G_n;\;\;\;V=U\in {\cal S}_n;\;\;\;W={\bf 1}_n\in {\cal S}_n.
\end{eqnarray}
Then
\begin{eqnarray}
{\mathfrak Q}=\frac{\bra{e}U\ket{f}}{g(\ket{f}\bra{e})};\;\;\;\theta \in G_n;\;\;\;V,W \in {\cal S}_n.
\end{eqnarray}
Since $\bra{g}U\ket{f}\le 1$ and $g(\ket{f}\bra{e})\ge 1$(Eq.(\ref{PO})) it follows that ${\mathfrak Q}\le 1$.

Many physically interesting quantities can be written as $\bra{e}U\ket{f}$ where $U$ is some unitary operator (e.g., Wigner and Weyl functions) and they lead to
 ${\mathfrak Q}\le 1$. 
In this sense the Grothendieck region $(1,k_G)$ is a new territory of Quantum Mechanics.

\end{example}

In this paper we are mainly interested in cases with ${\mathfrak Q}(\theta)$ in the Grothendieck region $(1,k_G)$.
We have seen earlier that for normal matrices $g(\Theta)\le ne_{\rm max}$.
We now point out, that
for normal matrices a necessary (but not sufficient) condition\cite{VOU1} for ${\mathfrak Q}>1$, is that $g(\Theta)<ne_{\rm max}$ (strict inequality) and that 
in the calculation of ${\mathfrak Q}(\theta)$ we should use $\theta=\lambda\Theta$ with
\begin{eqnarray}
\frac{1}{ne_{\rm max}}<\lambda<\frac{1}{g(\Theta)}.
\end{eqnarray}
If we use $\theta=\lambda\Theta$ with $\lambda\le \frac{1}{ne_{\rm max}}$, we get ${\mathfrak Q}(\theta)\le 1$.

\subsection{${\mathfrak Q}$ with the projectors $\Pi$ related to the coherent states }\label{sec57}

We prove the strict inequality $g(\Pi)<n$ for the $n\times n$ projector $\Pi$ in Eq.(\ref{proj}) (in this case $e_{\rm max}=1$).

\begin{proposition}
We consider the projector $\Pi$ in Eq.(\ref{proj}) corresponding to any of the coherent states ${\cal C}(d,n,z)$ described above.
For generic values of the parameter $z=\exp(i\theta)$, the strict inequality $g(\Pi)<n$ holds.
 \end{proposition}
\begin{proof}
We consider the columns $\vec{a}$, $\vec{b}$ that contain the elements $a_r$, $b_s$ correspondingly, with $|a_r|\le1$ and $|b_s|\le 1$ and
$r,s=0,...,n-1$. Then $\vec{\beta}=\Pi \vec{b}$ is the $n$-tuple representation of some vector in $H(d)$ (with respect to the coherent states involved in the definition of $\Pi$).
Indeed, $\vec{\beta}=\Pi \vec{b}=M^\dagger(M\vec{b})$ and therefore $\vec{\beta}$ is the $n$-tuple representation of $M\vec{b}\in H(d)$ (Eq.(\ref{60})).
Also $|\vec{a}|^2\le n$, $|\vec{b}|^2\le n$ and therefore $|\vec{\beta}|^2\le n$.
Then
\begin{eqnarray}
{\mathfrak C}(\Pi)=\left |\vec{a}\cdot (\Pi \vec{b})\right |= \left |\vec{a}\cdot\vec{\beta}\right |\le |\vec{a}||\vec{\beta}|;\;\;\;|\vec{a}|^2\le n;\;\;\;|\vec{\beta}|^2\le n.
\end{eqnarray}
For ${\mathfrak C}(\Pi)=n$ we need
$\vec{a}$ to be proportional to $\vec{\beta}$, and therefore both should be an $n$-tuple representation (not normalised) of some vector in $H(d)$.
In addition to that we  want {\bf all} $|a_r|$ to take their maximum value $1$, otherwise ${\mathfrak C}(\Pi)<n$.
But according to lemma \ref{L} this is not possible (for generic $z$ in the coherent states) . We conclude that all ${\mathfrak C}(\Pi)<n$, and therefore $g(\Pi)<n$.
\end{proof}

We now use Eq.(\ref{77}) with the three examples of coherent states discussed earlier.
The corresponding projectors lead to ${\mathfrak Q}>1$.
\begin{itemize}
\item
We consider the set of coherent states ${\cal C}(3,6,z)$, and the corresponding projector $\Pi$ in Eq.(\ref{P1}) which has ${\cal N}(\Pi)=\frac{1}{\sqrt{2}}$.
We use Eq.(\ref{77}) with $V=W=\sqrt{2}\Pi\in {\cal S}_6$.
In this case the strict inequality $g(\Pi)<6$ holds  (for generic $z$). We also take
\begin{eqnarray}
\theta=\lambda \Pi;\;\;\;\frac{1}{6}<\lambda<\frac{1}{g(\Pi)}.
\end{eqnarray}
and find ${\mathfrak Q}=|{\rm Tr}(\theta VW^\dagger )|=6\lambda>1$.
We note that  ${\mathfrak Q}$ is a measurable quantity. We consider the density matrix $\rho=\frac{1}{3}\Pi$ and the observable $\Psi=6\lambda\Pi$ and then ${\mathfrak Q}={\rm Tr}(\rho \Psi)$.

\item
We consider the set of coherent states ${\cal C}(4,8,z)$, and the corresponding projector $\Pi$ in Eq.(\ref{P2}) which has ${\cal N}(\Pi)=\frac{1}{\sqrt{2}}$.
We use Eq.(\ref{77}) with $V=W=\sqrt{2}\Pi\in {\cal S}_8$.
In this case the strict inequality $g(\Pi)<8$ holds. We also take
\begin{eqnarray}
\theta=\lambda \Pi;\;\;\;\frac{1}{8}<\lambda<\frac{1}{g(\Pi)}.
\end{eqnarray}
and find ${\mathfrak Q}=|{\rm Tr}(\theta VW^\dagger )|=8\lambda>1$.
 \item
 We consider the set of coherent states ${\cal C}(4,12,z)$, and the corresponding projector $\Pi$ in table \ref{t1} which has ${\cal N}(\Pi)=\frac{1}{\sqrt{3}}$.
 We use Eq.(\ref{77}) with $V=W=\sqrt{3}\Pi\in {\cal S}_{12}$. 
In this case the strict inequality $g(\Pi)<12$ holds (for generic $z$). We also take
\begin{eqnarray}
\theta=\lambda \Pi;\;\;\;\frac{1}{12}<\lambda<\frac{1}{g(\Pi)}.
\end{eqnarray}
and find ${\mathfrak Q}=|{\rm Tr}(\theta VW^\dagger )|=12\lambda>1$.
\end{itemize}

\section{The ultra-quantum coherent states violate logical Bell-like inequalities for a single quantum system }\label{sec58}

In this section we show that the three examples of coherent states discussed earlier, violate logical Bell-like inequalities
and in this sense they are far from quasi-classicality and they are deep into the quantum region.

Subsection \ref{S1} introduces briefly Frechet inequalities in a classical context, and shows that they hold for Kolmogorov probabilities but they do not hold for 
the more general concept of capacities.
They are based on Boole's inequality which holds for  Kolmogorov probabilities but not for capacities (non-additive probabilities).

Subsection \ref{S2} shows how quantum probabilities can be viewed as capacities.
Only for commuting projectors, the corresponding  probabilities obey relations analogous to those of Kolmogorov probabilities.

Subsection \ref{S3} presents logical Bell-like inequalities for a single quantum system.
Of course in a single quantum system we have no correlations between the various parties as in multipartite systems.
Physically, our inequalities resemble the Frechet probabilistic inequalities in a classical context.
But their mathematical derivation\cite{AH}, is the same for a single system or for multipartite systems.
The next three subsections show that they are violated by projectors related to coherent states in any of the orbits in the three sets
${\cal C}(3,6,z)$, ${\cal C}(4,8,z)$, ${\cal C}(4,12,z)$.

\subsection{Frechet inequalities for Kolmogorov probabilities and capacities}\label{S1}

Let $A_1,A_2$ be subsets of the set $\Omega$ of all alternatives.
Kolmogorov probabilities\cite{K} obey the relation
\begin{eqnarray}\label{K1}
p(A_1\cup A_2)+p(A_1\cap A_2)-p(A_1)-p(A_2)=0.
\end{eqnarray}
For exclusive sets, this gives the additivity relation
\begin{eqnarray}\label{2}
A_1\cap A_2=\emptyset\;\rightarrow\;p(A_1\cup A_2)=p(A_1)+p(A_2).
\end{eqnarray}
From this follows the `complement relation'
\begin{eqnarray}\label{41}
p(\overline A)=1- p(A).
\end{eqnarray}
Here ${\overline A}=\Omega\setminus A$ is the complement of the set $A$.
Also  the following Boole's inequality follows from Eq.(\ref{K1})
\begin{eqnarray}\label{44}
p(A_1\cup A_2)\le p(A_1)+p(A_2),
\end{eqnarray}
and  generalises into many $A_1,...,A_n$.

Capacities or non-additive probabilities replace Eq.(\ref{K1}) with  the monotonically increasing relation
\begin{eqnarray}\label{3}
A_1\subseteq A_2\;\rightarrow\;p(A_1)\le p(A_2),
\end{eqnarray}
which is weaker (Eq.(\ref{K1}) implies Eq(\ref{3}) but the opposite is not true).
In this case the additivity relation of Eq.(\ref{2}) does not hold in general.
Capacities have been introduced by Choquet \cite{C} and they have been used extensively in 
areas like artificial intelligence, operations research, mathematical economics,etc \cite{D1,D2,D3,D4,D5}.
In a quantum context they have been used in \cite{D6,D7,D8}.

For capacities
\begin{eqnarray}\label{L1}
p(A_1\cup A_2)+p(A_1\cap A_2)\ne p(A_1)+p(A_2),
\end{eqnarray}
in general. It follows that the additivity relation in Eq.(\ref{2}), Boole's inequality in Eq.(\ref{44}),  and the complement relation in Eq.(\ref{41}) do not hold in general.

\begin{proposition}\label{pro35}
Let $A_1,...,A_n$ be subsets of the set $\Omega$ such that $A_1\cap...\cap A_n=\emptyset$.
Then the following Frechet inequality\cite{F1,F2} holds for Kolmogorov probabilities (but not for capacities):
\begin{eqnarray}\label{4}
\sum_{i=1}^n p(A_i)\le (n-1).
\end{eqnarray}
\end{proposition}
\begin{proof}

We get
\begin{eqnarray}
A_1\cap...\cap A_n=\emptyset\;\rightarrow\;\overline A_1\cup...\cup\overline A_n=\Omega.
\end{eqnarray}
Using Boole's inequality we get
\begin{eqnarray}
p(\overline A_1)+...+p(\overline A_n)\ge p(\overline A_1\cup...\cup\overline A_n)=1.
\end{eqnarray}
Taking into account the complement relation in Eq.(\ref{41}) 
we prove Eq(\ref{4}).

Both Boole's inequality and the complement relation hold for Kolmogorov probabilities but not for capacities.
Therefore the proposition does not hold for capacities.
\end{proof}

Intuitively, the following inequality is very natural (because the subsets $A_i$ cover $\Omega$):
\begin{cor}
Let $A_1,...,A_n$ be subsets of the set $\Omega$ such that $A_1\cup...\cup A_n=\Omega$.
Then the following inequality holds for Kolmogorov probabilities (but not for capacities):
\begin{eqnarray}\label{12A}
\sum_{i=1}^n p(A_i)\ge 1.
\end{eqnarray}
\end{cor}
\begin{proof}
We get
\begin{eqnarray}
A_1\cup...\cup A_n=\Omega\;\rightarrow\;\overline A_1\cap...\cap\overline A_n=\emptyset.
\end{eqnarray}
Then proposition \ref{pro35} gives
\begin{eqnarray}
\sum_{i=1}^n p(\overline A_i)\le (n-1),
\end{eqnarray}
and using the complement relation in Eq.(\ref{41}) we prove Eq.(\ref{12A}).
The proof holds for Kolmogorov probabilities but not for capacities.
\end{proof}

\subsection{Quantum probabilities as capacities}\label{S2}

Let $\rho$ be a density matrix of a quantum system with Hilbert space $H(d)$ and $h\prec H(d)$. We use the following notation for quantum probabilities:
\begin{eqnarray}\label{30}
p(h|\rho)={\rm Tr}[\rho \Pi(h)];\;\;\;
p(h^\perp|\rho)=1-p(h|\rho).
\end{eqnarray} 
Clearly
\begin{eqnarray}
p[H(d)|\rho]=1;\;\;\;p({\cal O}|\rho)=0,
\end{eqnarray} 
and
\begin{eqnarray}\label{TT}
h_1\prec h_2\;\rightarrow\;p(h_1|\rho)\le p(h_2|\rho).
\end{eqnarray} 

Refs \cite{AH2,AH3} introduced the quantity
\begin{eqnarray}\label{33}
{\mathfrak D}(h_1,h_2)=\Pi(h_1\vee h_2)+\Pi(h_1\wedge h_2)-\Pi(h_1)-\Pi(h_2).
\end{eqnarray} 
Since the trace of a projector with a density matrix is a probability, this is a `correction' of  Eq(\ref{K1}) in a quantum context.
It has been shown that ${\mathfrak D}(h_1,h_2)$ is related to the commutator $[\Pi(h_1),\Pi(h_2)]$ as follows:
\begin{eqnarray}\label{DD}
[\Pi(h_1),\Pi(h_2)]={\mathfrak D}(h_1,h_2)[\Pi(h_1)-\Pi(h_2)].
\end{eqnarray} 
Also
\begin{eqnarray}
{\rm Tr}[{\mathfrak D}(h_1,h_2)]={\rm dim}(h_1\vee h_2)+{\rm dim}(h_1\wedge h_2)-{\rm dim}(h_1)-{\rm dim}(h_2)=0.
\end{eqnarray} 
We call this modularity relation, because the equality in the right hand side is an important property of any modular lattice.

For a density matrix $\rho$, Eq.(\ref{33}) gives the probabilistic relation
\begin{eqnarray}
{\rm Tr}[\rho\Pi(h_1\vee h_2)]+{\rm Tr}[\rho\Pi(h_1\wedge h_2)]-{\rm Tr}[\rho \Pi(h_1)]-{\rm Tr}[\rho\Pi(h_2)]={\rm Tr}[\rho{\mathfrak D}(h_1,h_2)].
\end{eqnarray} 
The right hand side can take positive or negative values because ${\mathfrak D}(h_1,h_2)$ has both negative and positive eigenvalues (its trace is zero).
Therefore quantum probabilities are capacities (they obey Eq.(\ref{TT})) and do not obey Boole's inequality, in general.

In the special case that the projectors $\Pi(h_1), \Pi(h_2)$ commute, Eq.(\ref{DD}) shows that ${\mathfrak D}(h_1,h_2)=0$.
In this special case, Boole's inequality holds.
In this sense, commutativity is related to `classical behaviour'.

\subsection{Logical Bell-like inequalities for a single quantum system: quantum versions of the Frechet probabilistic inequalities}\label{S3}

Using a quantum (Hilbert space) analogue of the derivation of the Frechet inequality, we prove logical Bell-like inequalities.
They have been introduced in ref.\cite{AH} and used in refs.\cite{AH1,AH2}, in the context of multipartite entangled systems. In this paper they are used with our coherent states in a single quantum system.
In the present context we have no correlations between the various parties, but we have probabilistic inequalities which resemble the Frechet probabilistic inequalities.

\begin{proposition}\label{pro12}
Let $h_1,...,h_n$ subspaces of $H(d)$ such that
\begin{eqnarray}\label{80}
h_1\wedge  ...\wedge h_n={\cal O}.
\end{eqnarray}
If Boole's inequality holds  (e.g., if the $\Pi(h_i)$ commute with each other) then for any density matrix $\rho$
\begin{eqnarray}\label{AA}
p(h_1|\rho)+...+p(h_n|\rho)\le n-1.
\end{eqnarray} 
\end{proposition}
\begin{proof}
\begin{eqnarray}
h_1^\perp\vee  ...\vee h_n^\perp=H(d)\;\rightarrow\;p(h_1^\perp\vee ...\vee h_n^\perp|\rho)=1.
\end{eqnarray} 
Assuming that Boole's inequality holds (e.g., if the $\Pi(h_i)$ commute with each other and then the $\Pi(h_i^\perp)$ also commute with each other), we get
\begin{eqnarray}\label{A}
p(h_1^\perp|\rho)+...+p(h_n^\perp|\rho)\ge p(h_1^\perp\vee  ...\vee h_n^\perp|\rho)=1,
\end{eqnarray} 
and using the fact that $p(h_i^\perp|\rho)=1-p(h_i|\rho)$ (Eq.(\ref{30})), we prove Eq.(\ref{AA}).
\end{proof}
Intuitively the following inequality is very natural (because the subspaces $h_i$ cover $H(d)$), and its violation by quantum mechanics is  counter-intuitive.
\begin{cor}
If
\begin{eqnarray}\label{89}
h_1\vee  ...\vee h_n=H(d),
\end{eqnarray} 
and Boole's inequality holds (e.g., if the $\Pi(h_i)$ commute with each other), then for a density matrix $\rho$
\begin{eqnarray}\label{AAA}
1\le p(h_1|\rho)+...+p(h_n|\rho).
\end{eqnarray} 
\end{cor}
\begin{proof}
The assumption in Eq.(\ref{89}) implies that
\begin{eqnarray}
h_1^\perp\wedge ...\wedge h_n^\perp={\cal O}
\end{eqnarray} 
and using proposition \ref{pro12} we get 
\begin{eqnarray}\label{A}
p(h_1^\perp|\rho)+...+p(h_n^\perp|\rho)\le n-1.
\end{eqnarray} 
From this and the fact that $p(h_i^\perp|\rho)=1-p(h_i|\rho)$ (Eq.(\ref{30})), we prove Eq.(\ref{AAA}).
\end{proof}
The proof in the above proposition and corollary is based on Boole's inequality.
We explained earlier that Boole's inequality holds for Kolmogorov probabilities and is violated (in general) by quantum  probabilities.
Consequently quantum probabilities might violate the inequalities in Eqs(\ref{AA}), (\ref{AAA}), and this is the case with our `ultra-quantum coherent states' (see below).

\subsection{Example with the coherent states in ${\cal C}(3,6,z)$}
 
 We use Eq.(\ref{AAD}) with the one-dimensional subspaces related to coherent states in the $0$-orbit and for generic $z$ we get (we use here the `single index notation'):
\begin{eqnarray}\label{HJ1}
h(0)\vee h(1)\vee h(2)=H(3).
\end{eqnarray}
 We use the notation
\begin{eqnarray}
&&\varpi (r)=\ket{a_z(r)}\bra{a_z(r)};\;\;\;\varpi (r)^\perp={\bf 1}_3-\varpi(r)\nonumber\\
&&p[h(r)|\rho]={\rm Tr}[\varpi (r)\rho];\;\;\;p[h(r)^\perp|\rho]={\rm Tr}[\varpi (r)^\perp \rho]=1-{\rm Tr}[\varpi (r)\rho].
\end{eqnarray}
From Eq.(\ref{HJ1}) follows that
\begin{eqnarray}
h(0)^\perp \wedge h(1)^\perp \wedge h(2)^\perp={\cal O}.
\end{eqnarray}
Then the inequality in Eq.(\ref{AA}) becomes
\begin{eqnarray}\label{5610}
p[h(0)^\perp|\rho]+p[h(1)^\perp|\rho]+p[h(2)^\perp|\rho]\le 2,
\end{eqnarray}
and the inequality in Eq.(\ref{AAA}) becomes
\begin{eqnarray}\label{57}
1\le p[h(0)|\rho]+p[h(1)|\rho]+p[h(2)|\rho].
\end{eqnarray}
Using Eq(\ref{35}) we get:
\begin{eqnarray}\label{52}
&&\varpi(0)+\varpi(1)+\varpi(2)=3\sigma_{00}={\bf 1}_3+A;\;\;\;A=\frac{1}{2}(z^*X+zX^\dagger)\nonumber\\
&&\varpi(0)^\perp+\varpi(1)^\perp+\varpi(2)^\perp=3\cdot{\bf 1}_3-3\sigma_{00}=2\cdot{\bf 1}_3-A
\end{eqnarray}
It follows that
\begin{eqnarray}\label{93}
&&p[h(0)^\perp|\rho]+p[h(1)^\perp|\rho]+p[h(2)^\perp|\rho]=2-{\rm Tr}(\rho A)\nonumber\\
&&p[h(0)|\rho]+p[h(1)|\rho]+p[h(2)|\rho]=1+{\rm Tr}(\rho A).
\end{eqnarray}
$A$ has both positive and negative eigenvalues (its trace is zero), and therefore we can find many examples with ${\rm Tr}(\rho A)<0$, in which case the logical Bell-like inequalities are violated.
Analogous results can be shown with projectors related to coherent states in the other orbit.

\subsection{Example with the coherent states in ${\cal C}(4,8,z)$}

We present analogous results to the above, with the coherent states in ${\cal C}(4,8,z)$.
We use Eq.(\ref{AAD}) with the spaces related to coherent states in the $0$-orbit, and for generic $z$ we get
\begin{eqnarray}\label{HJ2}
h(0)\vee h(1)\vee h(2)\vee h(3)= H(4).
\end{eqnarray}
From this  follows that
\begin{eqnarray}
h(0)^\perp \wedge h(1)^\perp \wedge h(2)^\perp\wedge h(3)^\perp={\cal O}.
\end{eqnarray}
The logical Bell-like inequalities are
\begin{eqnarray}\label{B1}
p[h(0)^\perp|\rho]+p[h(1)^\perp|\rho]+p[h(2)^\perp|\rho]+p[h(3)^\perp|\rho]\le 3,
\end{eqnarray}
and
\begin{eqnarray}\label{B2}
1\le p[h(0)|\rho]+p[h(1)|\rho]+p[h(2)|\rho]+p[h(3)|\rho].
\end{eqnarray}
Using Eq(\ref{45}) we get:
\begin{eqnarray}\label{52}
&&\varpi(0)+\varpi(1)+\varpi(2)+\varpi(3)=4\sigma_{00}={\bf 1}_4+A;\;\;\;A=\frac{1}{2}(z^*X+zX^\dagger)\nonumber\\
&&\varpi(0)^\perp+\varpi(1)^\perp+\varpi(2)^\perp+\varpi(3)^\perp=4\cdot{\bf 1}_4-4\sigma_{00}=3\cdot{\bf 1}_4-A.
\end{eqnarray}
It follows that
\begin{eqnarray}\label{93}
&&p[h(0)^\perp|\rho]+p[h(1)^\perp|\rho]+p[h(2)^\perp|\rho]+p[h(3)^\perp|\rho]=3-{\rm Tr}(\rho A)\nonumber\\
&&p[h(0)|\rho]+p[h(1)|\rho]+p[h(2)|\rho]+p[h(3)|\rho]=1+{\rm Tr}(\rho A).
\end{eqnarray}
$A$ has both positive and negative eigenvalues (its trace is zero), and therefore we can find many examples with ${\rm Tr}(\rho A)<0$, in which case the logical Bell-like inequalities are violated.
Analogous results can be shown with projectors related to coherent states in the other orbit.

\subsection{Example with the coherent states in ${\cal C}(4,12,z)$}
As in the previous sections, we use Eq.(\ref{AAD}) with the spaces related to coherent states in the $0$-orbit in ${\cal C}(4,12,z)$, and for generic $z$ we get
\begin{eqnarray}\label{HJ3}
h(0)\vee h(1)\vee h(2)\vee h(3)=H(4).
\end{eqnarray}
From this follows that
\begin{eqnarray}
h(0)^\perp \wedge h(1)^\perp \wedge h(2)^\perp\wedge h(3)^\perp={\cal O}.
\end{eqnarray}

The logical Bell-like inequalities are
\begin{eqnarray}\label{CV1}
p[h(0)^\perp|\rho]+p[h(1)^\perp|\rho]+p[h(2)^\perp|\rho]+p[h(3)^\perp|\rho]\le 3,
\end{eqnarray}
and
\begin{eqnarray}\label{CV2}
1\le p[h(0)|\rho]+p[h(1)|\rho]+p[h(2)|\rho]+p[h(3)|\rho].
\end{eqnarray}

Using Eq(\ref{GG3}) we get
\begin{eqnarray}\label{52}
&&\varpi(0)+\varpi(1)+\varpi(2)+\varpi(3)=4\sigma_{00}={\bf 1}_4+A\nonumber\\
&&\varpi(0)^\perp+\varpi(1)^\perp+\varpi(2)^\perp+\varpi(3)^\perp=4\cdot {\bf 1}_4-4\sigma_{00}=3\cdot {\bf 1}_4-A\nonumber\\
&&A=\frac{1}{3}(z^*+1)X^3+\frac{1}{3}(z+z^*)X^2+\frac{1}{3}(z+1)X
\end{eqnarray}
Therefore
\begin{eqnarray}\label{93}
&&p[h(0)^\perp|\rho]+p[h(1)^\perp|\rho]+p[h(2)^\perp|\rho]+p[h(3)^\perp|\rho]=3-{\rm Tr}(\rho A)\nonumber\\
&&p[h(0)|\rho]+p[h(1)|\rho]+p[h(2)|\rho]+p[h(3)|\rho]=1+{\rm Tr}(\rho A)
\end{eqnarray}
As in the previous examples, $A$ has both positive and negative eigenvalues (its trace is zero), and therefore we can find many examples with ${\rm Tr}(\rho A)<0$, in which case the logical Bell-like inequalities are violated.

\section{Generalisations}\label{sec7}
Above we gave three families of examples with states that have all the properties {\bf C1-C6}.
We can generalise these examples to spaces with larger dimension, 
and below we give three such examples.
The states in these examples have the coherence properties {\bf C1-C4},
but it is an open problem whether they also have the `ultra-quantum properties' {\bf C5} (${\mathfrak Q}>1$) and {\bf C6} which are at the heart of this paper.

The general approach is to find a $d\times n$ matrix $M$ such that
\begin{eqnarray}\label{PROJ}
MM^\dagger={\bf 1}_d;\;\;\;M^\dagger M=\Pi;\;\;\;\Pi^2=\Pi.
\end{eqnarray}
$MM^\dagger={\bf 1}_d$ is the resolution of the identity for the coherent states (columns).
$\Pi$ is the projector of the overlaps of the coherent states (Eq.(\ref{proj})).
The matrix $M$ consists of $\frac{n}{d}$ blocks each of which contains the columns $X^r\ket{v}$.
So action of the cyclic group ${\cal G}_d$ leads to $\frac{n}{d}$ orbits.

\subsection{The set ${\cal C}(5,10,z)$}
An example in $H(5)$ is the set ${\cal C}(5,10,z)$ that contains the (normalized) columns of
\begin{eqnarray}
&&M=\frac{1}{2}
\setcounter{MaxMatrixCols}{10}
\begin{pmatrix}
z&1&0&0&0&-z&1&0&0&0\\
1&0&0&0&z&1&0&0&0&-z\\
0&0&0&z&1&0&0&0&-z&1\\
0&0&z&1&0&0&0&-z&1&0\\
0&z&1&0&0&0&-z&1&0&0
\end{pmatrix};\;\;\;|z|=1.
\end{eqnarray}
We can check that Eq.(\ref{PROJ}) holds.
We can also check that all the details of properties {\bf C2,C3,C4} hold,
but it is an open problem whether they obey properties {\bf C5} (${\mathfrak Q}>1$) and {\bf C6}.

\subsection{The set ${\cal C}(5,15,z)$}
Another example in $H(5)$ is the set ${\cal C}(5,15,z)$ that contains the (normalized) columns of
\begin{eqnarray}
&&M=\frac{1}{3}
\setcounter{MaxMatrixCols}{15}
\begin{pmatrix}
z&1&1&0&0&z&\omega&\omega^2&0&0&z&\omega ^2& \omega&0&0\\
1&1&0&0&z&\omega&\omega^2&0&0&z&\omega^2&\omega&0&0&z\\
1&0&0&z&1&\omega^2&0&0&z&\omega&\omega&0&0&z&\omega^2\\
0&0&z&1&1&0&0&z&\omega&\omega^2&0&0&z&\omega^2&\omega\\
0&z&1&1&0&0&z&\omega&\omega^2&0&0&z&\omega^2&\omega&0
\end{pmatrix};\;\;\;\omega=\exp\left (i\frac{2\pi}{3}\right );\;\;\;|z|=1.
\end{eqnarray}
We can check that Eq.(\ref{PROJ}) holds.
We can also check that all the details of properties {\bf C2,C3,C4} hold,
but it is an open problem whether they obey properties {\bf C5} (${\mathfrak Q}>1$) and {\bf C6}.

\subsection{The set ${\cal C}(6,12,z)$}
An example in $H(6)$ is the set ${\cal C}(6,12,z)$ that contains the (normalized) columns of
\begin{eqnarray}
&&M=\frac{1}{2}
\setcounter{MaxMatrixCols}{12}
\begin{pmatrix}
z&1&0&0&0&0&-z&1&0&0&0&0\\
1&0&0&0&0&z&1&0&0&0&0&-z\\
0&0&0&0&z&1&0&0&0&0&-z&1\\
0&0&0&z&1&0&0&0&0&-z&1&0\\
0&0&z&1&0&0&0&0&-z&1&0&0\\
0&z&1&0&0&0&0&-z&1&0&0&0\\
\end{pmatrix};\;\;\;|z|=1.
\end{eqnarray}
We can check that Eq.(\ref{PROJ}) holds.
We can also check that all the details of properties {\bf C2,C3,C4} hold,
but it is an open problem whether they obey properties {\bf C5} (${\mathfrak Q}>1$) and {\bf C6}.

\section{Discussion}
There is much work in the literature on the classical-quantum interface (semiclassical limit).
The present paper is in the other end of the spectrum (the `ultra-quantum limit').
It uses the Grothendieck formalism which replaces scalar quantities in ${\mathfrak C}$ with vectors in ${\mathfrak Q}$, and finds that when 
${\mathfrak C}\in (0,1)$ then ${\mathfrak Q}\in (0,k_G)$. 
The `Grothendieck region' $(1,k_G)$ is classically forbidden in the sense that  ${\mathfrak C}$ does not take values in it, and it can be viewed as `ultra-quantum region'.

In the context of a single quantum system with a $d$-dimensional Hilbert space $H(d)$, we have introduced novel quantum states with two sets of properties:
\begin{itemize}
\item
{\bf Coherence properties} ({\bf C1}-{\bf C4}):
Resolution of the identity, invariance under transformations in a finite cyclic group ${\cal G}_d$, discrete isotropy, and other discrete symmetries.
The action of the cyclic group ${\cal G}_d$ partitions the set of coherent states ${\cal C}(d,n,z)$ into $\frac{n}{d}$ orbits.
 A $n$-tuple representation of states in $H(d)$ is the discrete analogue of the Bargmann analytic representation.
Examples of these coherent states are the sets ${\cal C}(3,6,z)$, ${\cal C}(4,8,z)$, ${\cal C}(4,12,z)$ in section \ref{sec31} and also the sets
${\cal C}(5,10,z)$, ${\cal C}(5,15,z)$, ${\cal C}(6,12,z)$ in section \ref{sec7}. 
\item
{\bf Ultra-quantum properties} ({\bf C5},{\bf C6}):
The main `ultra-quantum' property is related to the Grothendieck formalism (which is used here in the context of a single quantum system).
Examples that give ${\mathfrak Q}$ in the Grothendieck region $(1,k_G)$ are at the `edge' of the Hilbert space and quantum formalisms.
We have seen in example \ref{ex1} that large families of physical quantities  lead to ${\mathfrak Q}\le 1$, and in this sense examples with ${\mathfrak Q}> 1$ are
rare. We have shown that projectors related to our coherent states in the sets ${\cal C}(3,6,z)$, ${\cal C}(4,8,z)$, ${\cal C}(4,12,z)$ in section \ref{sec31}, lead to ${\mathfrak Q}$ in the Grothendieck region $(1,k_G)$.

Furthermore our coherent states violate  logical Bell-like inequalities, which for a single quantum system are similar to Frechet probabilistic inequalities.
These inequalities are based on Boole's inequality which holds for Kolmogorov probabilities and is violated by quantum probabilities.
The violation of these inequalities by our coherent states, shows that they are deep into the quantum region.

\end{itemize}

We conclude with two open problems:
\begin{itemize}
\item
The projectors used with the ultra-quantum properties  {\bf C5} (${\mathfrak Q}>1$) and {\bf C6} are reproducing kernels.
One of the main applications of reproducing kernels is in maximisation problems in areas like machine learning (e.g., \cite{A1,A2,A3}).
So the reproducing kernel property of the projectors, might provide a deeper insight into why they give high values of ${\mathfrak Q}$.
\item
It is an open problem whether the states in the sets ${\cal C}(5,10,z)$, ${\cal C}(5,15,z)$, ${\cal C}(6,12,z)$ in section \ref{sec7}, have the ultra-quantum properties ({\bf C5},{\bf C6}).
\end{itemize}

In summary, the work studies novel coherent states which are deep into the quantum region.

\newpage

\begin{table} 
\small
\caption{ The projector $\Pi_{rs}=\frac{1}{3}\bra{a_z(r)}a_z(s)\rangle $ where ${\ket{a_z(r)}}$ are the columns in the matrix $M$ in Eq.(\ref{AQ3}) (normalised).
The matrix elements of $9\Pi_{rs}$  are shown. Here $\omega=\exp(i\frac{2\pi}{3})$ and $|z|=1$. }
\def\arraystretch{2}
\scalebox{0.8}{
\begin{tabular}{|c|c|c|c|c|c|c|c|c|c|c|c|}\hline
$3$&$z^*+1$&$z^*+z$&$z+1$&$0$&$z^*\omega+\omega^2$&$z^*\omega^2+z$&$z+\omega$&$0$&$z^*\omega^2+\omega$&$z^*\omega+z$&$z+\omega^2$\\\hline
$z+1$&$3$&$1+z^*$&$z+z^*$&$z+\omega$&$0$&$\omega^2+z^*\omega$&$z+\omega^2z^*$&$z+\omega^2$&$0$&$\omega+z^*\omega^2$&$z+z^*\omega$\\\hline
$z+z^*$&$1+z$&$3$&$z^*+1$&$z+z^*\omega ^2$&$\omega+ z$&$0$&$z^*\omega+\omega^2$&$z+z^*\omega$&$\omega^2+z$&$0$&$z^*\omega^2+\omega$\\\hline
$z^*+1$&$z^*+z$&$z+1$&$3$&$z^*\omega+\omega^2$&$z^*\omega^2+z$&$z+\omega$&$0$&$z^*\omega^2+\omega$&$z^*\omega+z$&$z+\omega^2$&$0$\\\hline
$0$&$z^*+\omega^2$&$z^*+\omega z$&$z\omega^2+\omega$&$3$&$z^*\omega+\omega$&$z^*\omega^2+z\omega$&$z \omega^2+\omega^2$&$0$&$z^*\omega^2+1$&$z^*\omega+z\omega$&$z\omega^2+1$\\\hline
$z\omega^2+\omega$&$0$&$\omega^2+z^*$&$z\omega+z^*$&$z\omega^2+\omega^2$&$3$&$\omega+ z^*\omega$&$z\omega +z^*\omega^2$&$z\omega^2+1$&$0$&$1+z^*\omega^2$&$z^*\omega+z\omega$\\\hline
$z\omega+z^*$&$\omega+z\omega^2$&$0$&$z^*+\omega^2$&$z\omega+ z^*\omega^2$&$\omega ^2z+\omega^2$&$3$&$z^*\omega+\omega$&$z\omega+z^*\omega$&$z\omega^2+1$&$0$&$z^*\omega^2+1$\\\hline
$\omega ^2+z^*$&$\omega z+z^*$&$\omega^2 z+\omega $&$0$&$z^*\omega+\omega$&$z^* \omega^2+z\omega$&$z\omega^2+\omega^2$&$3$&$z^*\omega^2+1$&$z^*\omega+z\omega$&$z\omega^2+1$&$0$\\\hline
$0$&$z^*+\omega$&$\omega^2 z+z^*$&$z\omega+\omega^2$&$0$&$z^*\omega+1$&$z^*\omega^2+z\omega^2$&$z\omega+1$&$3$&$z^*\omega^2+\omega^2$&$z^*\omega+z\omega^2$&$z\omega+\omega$\\\hline
$z\omega+\omega^2$&$0$&$z^*+\omega$&$z\omega^2+z^*$&$z\omega+1$&$0$&$1+z^*\omega$&$z\omega^2+z^*\omega^2$&$z\omega+\omega$&$3$&$\omega^2+z^*\omega^2$&$z\omega^2+z^*\omega$\\\hline
$z\omega^2+z^*$&$z\omega+\omega^2$&$0$&$z^*+\omega$&$z\omega^2+z^*\omega^2$&$z\omega+1$&$0$&$1+z^*\omega$&$z\omega^2+z^*\omega$&$\omega+z\omega$&$3$&$z^*\omega^2+\omega^2$\\\hline
$z^*+\omega$&$z^*+z\omega^2$&$z\omega+\omega^2$&$0$&$z^*\omega+1$&$z^*\omega^2+z\omega^2$&$z\omega+1$&$0$&$z^*\omega^2+\omega^2$&$z^*\omega+z\omega^2$&$z\omega+\omega$&$3$\\\hline
\end{tabular} \label{t1}
}
\end{table}

\end{document}